\documentclass[aps,prb,onecolumn,nofootinbib,citeautoscript,10pt]{revtex4-2}

\usepackage{amsmath,amssymb} 
\usepackage{comment}

\usepackage[tight]{subfigure} 

\usepackage[dvipsnames]{xcolor} 
\usepackage[papersize={8.5in,11in}]{geometry}
\usepackage[colorlinks=true]{hyperref}
\hypersetup{
    bookmarks=true,         
    unicode=false,          
    pdftoolbar=true,        
    pdfmenubar=true,        
    pdffitwindow=false,     
    pdfstartview={FitH},    
    pdfkeywords={keyword1} {key2} {key3}, 
    pdfnewwindow=true,      
    colorlinks=true,       
    linkcolor=magenta, 
    citecolor=blue,        
    filecolor=magenta,      
    urlcolor=blue           
} 

\usepackage{dcolumn}
\usepackage{color}
\usepackage{amssymb,amsmath}
\usepackage{tabularx,graphicx}
\usepackage{epstopdf}
\usepackage{latexsym}
\usepackage{colortbl}
\usepackage{psfrag}
\usepackage{bbm,bm,array,physics}
\usepackage{dsfont}
\usepackage{float, mathrsfs}

\def \nn{\nonumber \\}
\def\*#1{\mathbf{#1}} 

\begin{document}

\title{Reflections of topological properties in the planar-Hall response for semimetals carrying pseudospin-1 quantum numbers}

\author{Firdous Haidar}
\author{Ipsita Mandal}
\email{ipsita.mandal@snu.edu.in}

\affiliation{Department of Physics, Shiv Nadar Institution of Eminence (SNIoE), Gautam Buddha Nagar, Uttar Pradesh 201314, India}

\begin{abstract} 
We continue our investigations of the nature of the linear-response tensors in planar-Hall and planar-thermal Hall configurations, involving three-dimensional nodal-point semimetals, by considering here nodes hosting pseudospin-1 quasiparticles. Such systems exemplify multifold semimetals, as they have three bands crossing at a nodal point. We derive the explicit expressions of the electric, thermoelectric, and thermal coefficients, when the nodes are subjected to the combined influence of an electric field (and/or temperature gradient) and a weak (i.e., nonquantizing) magnetic field. In order to have a complete description, we consider the effects of the Berry curvature and the orbital magnetic moment on an equal footing, both of which originate from the underlying topological features of the bandstructure. Going beyond our previous works, we determine the out-of-plane response comprising the intrinsic anomalous-Hall and the Lorentz-force-contributed currents, and chalk out the effects of internode scatterings as well. Our theoretical explorations shed light on the mechanisms of transport in multifold semimetals, which are being investigated in contemporary experiments.
\end{abstract}

\maketitle
\tableofcontents

\section{Introduction}

There has been an immense amount of interest, comprising both theoretical and experimental efforts, for discovering and understanding novel transport characteristics shown by three-dimensional (3d) semimetals with symmetry-protected band-crossings. Such interest is triggered by the fact that, when a nodal point lies close to the Fermi level, the exotic signatures of transport-properties (such as the linear-response coefficients) often offer a glimpse of the mathematical concepts of topology being realized in the 3d Brillouin zone (BZ) of solid-state systems~\cite{burkov11_Weyl, yan17_topological, ips-kush-review, bernevig,ips-biref, grushin-multifold, ips-hermann-review, claudia-multifold}. Due to the fact that the density-of-states vanishes identically at the nodal points, the semimetallic bands differ in their behaviour from both the insulators (characterized by a finite gap between the bands) and the conventional metals (characterized by a overlap in finite regions of the BZ and, thus, featuring a finite density-of-states).
The usual semi-empirical approach of deriving the so-called $\boldsymbol k \cdot \boldsymbol p $ Hamiltonian, for visualizing the associated bandstructure, furnishes the low-energy effective Hamiltonian in the vicinity of a band-crossing point. For the case when we have $(2\, \varsigma + 1) $ bands crossing at the point, in the 3d momentum space parametrized by $\boldsymbol k  =\lbrace k_x, k_y, k_z  \rbrace $, the Hamiltonian takes the form of
$
\boldsymbol d( \boldsymbol{k} ) \cdot \boldsymbol {\mathcal S } $, where
$\boldsymbol d( \boldsymbol{k} ) = 
\lbrace d_x ( \boldsymbol{k} ) , d_y ( \boldsymbol{k} ) , d_z ( \boldsymbol{k} )  
 \rbrace.$
Here, $\boldsymbol {\mathcal S } $ is the vector operator comprising the components $ \lbrace \mathcal S_x, \mathcal S_y, \mathcal S_z  \rbrace $, representing the three components of the angular momentum operator in the spin-$\varsigma$ representation of the SU(2) group. Therefore, an appropriate set of of three $(2\, \varsigma + 1)  \times (2\, \varsigma + 1)$ square matrices constitute a bonafide representation, reproducing the $(2\, \varsigma + 1)$-band system, with the bands carrying the azimuthal quantum numbers spanning from $- \varsigma $ to $ \varsigma $. Consequently, the energy levels are labelled with these \textit{pseudospin} quantum numbers, invoking their analogy with the relativistic \textit{spin} quantum numbers. While the latter set arises from the spatial rotations of the Lorentz group, representing the fundamental spacetime symmetries, the former set is the consequence of (nonrelativistic) crystal symmetries. One intriguing outcome is that integer-valued pseudospin quantum numbers are possible, corresponding to an odd number of fermionic bands crossing at a nodal point, in contrast with the fact that integer values of spin are forbidden for relativistic fermionic particles (as embodied in the spin-statistics theorem). 

The poster-child of 3d semimetals is the pseudospin-1/2 Weyl semimetal (WSM)~\cite{burkov11_Weyl, armitage_review, yan17_topological}, featuring the simplest case of twofold band-crossing points, with a linear-in-momentum dispersion. Making things more exciting, multifold band-crossings have
been discovered in the 65 chiral space groups comprising chiral crystals~\cite{grushin-multifold}, which have only orientation-preserving symmetries. Such examples include pseudospin-1 triple-point semimetal (TSM)~\cite{bernevig, ips3by2, ady-spin1, krish-spin1, ips-cd1, tang2017_multiple, grushin-multifold, prb108035428, ips-abs-spin1, claudia-multifold} and pseudospin-3/2 Rarita-Schwinger-Weyl (RSW) semimetal~\cite{bernevig, long, igor, igor2, isobe-fu, tang2017_multiple, ips3by2, ips-cd1, ma2021_observation, ips-magnus, ips-jns, prb108035428, ips_jj_rsw, grushin-multifold, claudia-multifold, ips-rsw-ph, ips-shreya}, which have
three and four bands crossing at the nodal point, respectively.

The intensive interest in studying nodal-point semimetals primarily stems from the fact that their BZ harbours a nontrivial topology, with the nodes being the singular points of the vector-field lines of the Berry curvature (BC). The BC arises from the Berry phase, which endows the BZ with a nontrivial topology \cite{xiao_review, sundaram99_wavepacket, timm, ips_rahul_ph_strain, graf-Nband, rahul-jpcm, ips-kush-review, claudia-multifold, ips-ruiz, ips-rsw-ph, ips-tilted, ips-shreya} (when we visualize the BZ as a closed 3d manifold). The Berry phase also gives rise to the orbital magnetic moment (OMM) as another intrinsic topological property. Such topological features manifest themselves in various transport measurements, e.g., intrinsic anomalous-Hall effect~\cite{haldane04_berry,goswami13_axionic, burkov14_anomolous}, nonzero planar-Hall response \cite{zhang16_linear, chen16_thermoelectric, nandy_2017_chiral, nandy18_Berry, amit_magneto, das20_thermal, das22_nonlinear, pal22a_berry, pal22b_berry, fu22_thermoelectric, araki20_magnetic, mizuta14_contribution, ips-serena, ips_rahul_ph_strain, timm, rahul-jpcm, ips-kush-review, claudia-multifold, ips-ruiz, lei_li-ph, ips-tilted, ips-rsw-ph, ips-shreya}, magneto-optical conductivity under quantizing magnetic fields~\cite{gusynin06_magneto, staalhammar20_magneto, yadav23_magneto}, Magnus Hall effect~\cite{papaj_magnus, amit-magnus, ips-magnus}, circular dichroism \cite{ips-cd1, ips_cd}, circular photogalvanic effect \cite{moore18_optical, guo23_light,kozii, ips_cpge}, and transmission of quasiparticles across potential barriers/wells \cite{ips_aritra, ips-sandip, ips-sandip-sajid, ips-jns}.

The nontriviality of topology is manifested by nonzero BC monopoles \cite{fuchs-review, polash-review}, sitting at the nodal points, serving as topological charges, and sourcing the BC flux. These are equivalent to the Chern numbers, using the terminology from topology. The sign of the BC-monopole's charge is demarcated as the chirality $\chi$ of the node, leading to the notion of \textit{chiral} quasiparticles. They are said to be \textit{right-handed} or \textit{left-handed}, depending on whether $\chi = 1$ or $\chi = -1$. Summing over of all the topological charges in the BZ, carried either by the conduction or the valence bands of all the chirally-charged nodes, must give zero in a system resulting from electrons hopping on a crystal lattice. This is explained mathematically by the Nielsen-Ninomiya theorem \cite{nielsen81_no}. In bandstructures, this is reflected by the fact that we must have conjugate pairs of nodes in the BZ, having $\chi = \pm1$. The conventional way is to assign $\chi$ the sign of the monopole-charges of the negative-energy bands, where positive or negative is measured with respect to the band-touching point (assigned the zero of energy). Also, according to this convention, a positive (negative) sign is associated with a node taken to act as a source (sink) for BC-flux lines.

\begin{figure*}[t]
\subfigure[]{\includegraphics[width=0.225 \linewidth]{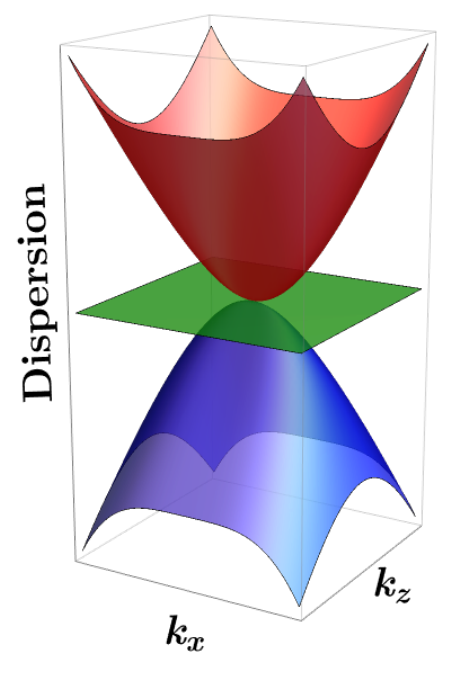}} 
\hspace{2 cm}
\subfigure[] {\includegraphics[width=0.475 \linewidth]{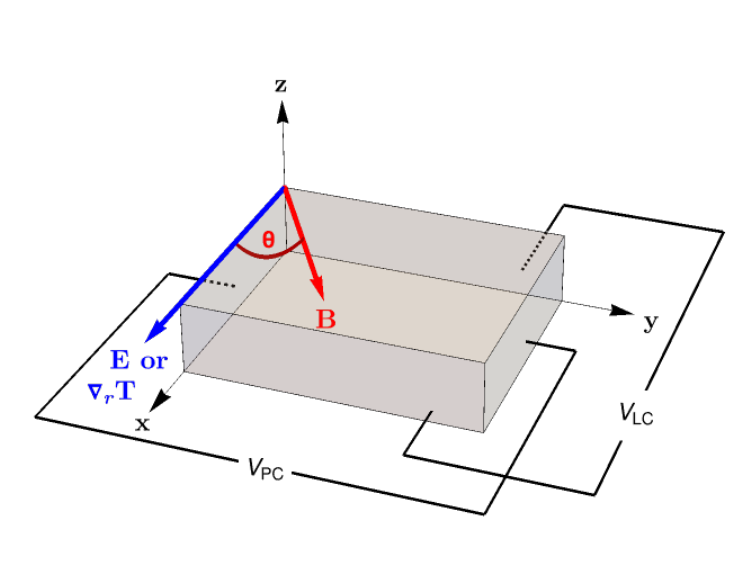}}
\caption{\label{figdis}(a) The dispersion of a single pseudospin-1 node, with $J=2 $, against the $k_z k_x$-plane.
(b) Schematics of the planar-Hall set-up where a semimetallic material, hosting pseudospin-$1$ quasiparticles, is probed with a static electric field $ \boldsymbol{E} = E\, {\boldsymbol{\hat x}} $ (and/or temperature gradient $ \nabla_{\boldsymbol r} T = \partial_x T\, {\boldsymbol{\hat x}}$), under the actional of a nonquantizing magnetic field $\boldsymbol B $. The latter makes an angle $\theta $ with respect to the electric field (and/or the temperature gradient). The in-plane voltages, generated parallel and perpendicular to $ E\, {\mathbf{\hat x}} $ (or $\partial_x T\, {\mathbf{\hat x}}$), are indicated by $V_{\rm LC}$ and $V_{\rm PC}$, respectively. The subscripts indicate their association with the longitudinal and transverse (i.e., planar-Hall) components of the resulting currents.}
\end{figure*}

In this paper, we focus on 3d TSMs, carrying the pseudospin value of $ \varsigma = 1$, whose effective low-energy continuum Hamiltonian is of the form of $ \boldsymbol d (\boldsymbol{k} )\cdot \boldsymbol  S  $ [cf. Fig.~\ref{figdis}(a)]. Here, $ \boldsymbol  S  $ represents the vector of the three matrices forming a spin-1 representation of the SO(3) group. The pseudospin-1 quasiparticles can be realized in various systems and have been studied extensively in wide contexts \cite{optical_lat1, optical_lat2, cold-atom, lv, spin13d1, spin13d2, spin13d3, spin13d4, shen,lan, urban, peng-he, lai, ips3by2, krish-spin1, ips-cd1, bitan-spin1, pal22b_berry, ips-abs-spin1}. They are often dubbed as ``Maxwell fermions'' \cite{cold-atom}, reflecting the analogy with the spin-1 quantum number of the photons (which are described by the Maxwell equations). When the effective low-energy dispersion of the quasiparticles, in the vicinity of a node, goes as $\sim \sqrt{\alpha_J^2 \, k_\perp^{2J} + v_z^2 \, k_z^2}$ (where $J \in \lbrace 1, 2, 3  \rbrace$, $\alpha_J $ is a material-dependent parameter, and $k_\perp = \sqrt {k_x^2 +k_y^2}$), it hosts a Chern number of magnitude $2J$ \cite{krish-spin1, bitan-spin1, pal22b_berry}. Therefore, for $J>1$, they form the analogues of the pseudospin-1/2 multi-Weyl semimetals (mWSMs) \cite{bernevig, bernevig2, ips_rahul_ph_strain, rahul-jpcm, ips-ruiz, ips-tilted}, differing in the number of bands and the value of the Chern number/monopole-charge (by a factor of two).

Here, we continue our explorations of planar-Hall and planar-thermal-Hall set-ups, comprising a semimetal subjected to the combined influence of an electric field, $\boldsymbol{E}$, (and/or a temperature gradient, $\nabla_{\boldsymbol{r}} T$) and a magnetic field, $\boldsymbol{B} $, as depicted in Fig.~\ref{figdis}(b). As pointed out above, we will consider the case when the semimetal harbours pseudospin-1 quasiparticles. In recent times, there have been extensive amounts of theoretical and experimental investigations involving the associated response coefficients~\cite{zhang16_linear, chen16_thermoelectric, nandy_2017_chiral, nandy18_Berry, amit_magneto, das20_thermal, das22_nonlinear, pal22a_berry, pal22b_berry, fu22_thermoelectric, araki20_magnetic, mizuta14_contribution, ips-serena, timm, onofre, ips_rahul_ph_strain, rahul-jpcm, ips-kush-review, claudia-multifold, ips-ruiz, ips-tilted, ips-internode}. The plane containing $\boldsymbol E$ (or $\nabla_{\boldsymbol{r}} T$) and $\boldsymbol B $ is chosen to be the $xy$-plane, with $ \boldsymbol E $ (or $\nabla_{\boldsymbol r} T$) fixed along the $x$-axis. The direction of the magnetic field, $\boldsymbol{B} \equiv  B \left( \cos \theta \, \hat{\boldsymbol{x}} +
\sin \theta\, \hat{\boldsymbol{y}} \right )$ (where $B \equiv \vert \boldsymbol{B}\vert $), is specified by the angle $\theta $. Clearly, $\theta $ is the angle it makes with $\hat{\boldsymbol{x}}$, and is not necessarily equal to $ \pi/ 2$ or $3\,\pi/2$ in general. We focus on the linear-response regimes with respect to the probe fields of $\boldsymbol E $ and $\nabla_{\boldsymbol r} T$. The relevant response tensors are the ones relating the electric and thermal currents to $\boldsymbol E $ and $\nabla_{\boldsymbol r} T$, which encompass the magnetoelectric conductivity ($\sigma_\chi$), the magnetothermoelectric conductivity ($\alpha_\chi $), and the magnetothermal conductivity ($\kappa_\chi$). Let $\ell_\chi $ denote the tensor relating the heat current to the temperature gradient at a vanishing electric field. Knowing $\ell_\chi $, one can compute $\kappa_\chi$ --- hence, $\ell_\chi $ itself is often loosely referred to as the magnetothermal coefficient. In summary, if we know the three independent tensors embodied by $\sigma_\chi$, $\alpha_\chi $, and $\ell_\chi $, we can construct all the response characteristics in transport measurements.
It is worth mentioning that the nature of $\sigma_\chi $ and $\ell_\chi $ has been studied in Ref.~\cite{pal22b_berry} for the case of $J=2$, but without taking the OMM-induced contributions into account.

The paper is organized as follows: In Sec.~\ref{secmodel}, we describe the explicit form of the low-energy effective Hamiltonian in the vicinity of a TSM node, for the three possible values of $J$. The resulting expressions for the BC and the OMM are also shown. Secs.~\ref{secsigma} and \ref{secalpha} are devoted to demonstrating the explicit expressions for the longitudinal and transverse components of $\sigma_\chi $, $\alpha_\chi$, and $\ell_\chi$, respectively.
In Sec.~\ref{secinter}, we discuss the effects of internode scatterings. Finally, we conclude with a summary and outlook in Sec.~\ref{secsum}. The appendices are devoted to elaborating on much of the details of the intermediate steps, necessary to derive the final expressions shown in the main text. 
In all our expressions, we will be using the natural units, which means that the reduced Planck's constant ($\hbar $), the speed of light ($c$), and the Boltzmann constant ($k_B $) are each set to unity.
Additionally, electric charge has no units, with the magnitude of a single electronic charge measuring $e =1$.

\section{Model}
\label{secmodel}

 
For the TSMs, which are characterized by nodal points with threefold degeneracies, we encounter two distinct situations that may arise to satisfy the Nielsen-Ninomiya theorem: Existence of a pair of conjugate nodes (1) of the same pseudospin variety, with $\chi = \pm 1$; (2) comprising bands of different pseudospin quantum numbers. The first case is exemplified by a pair of TSMS \cite{bernevig, ady-spin1}. The second possibility is exemplified by the following two cases:
\\(a) A single node of TSM is pinned at the centre of the BZ (i.e., the $\Gamma $-point), carrying a monopole charge of $+ \, 2 $, while a fourfold-degenerate node (comprising two copies of WSMs of the same chirality) exists at the boundary of the BZ (i.e., the $R$-point) with a net monopole charge equalling $- \,1- 1 = - \, 2$. Candidate materials include CoSi \cite{claudia-multifold}.
\\(b) In a typical material harbouring an RSW node \cite{tang2017_multiple, prb108035428, prl119206401, yamakage}, we find that there is the RSW node at the $\Gamma $-point carrying $+ \, 4 $ charge, and a sixfold-degenerate (originating from the doubling of pseudospin-1 excitations) at the $ R $-point carrying $- \, 4 $ charge. Candidate materials include the SrGePt family (e.g., SrSiPd, BaSiPd, CaSiPt, SrSiPt, BaSiPt, and BaGePt) \cite{prb108035428}.

Expanding the $\boldsymbol{k} \cdot \boldsymbol {p}$ Hamiltonian about a threefold-degeneracy point in small $\lbrace k_x, \, k_y , \, k_z \rbrace $, we obtain the low-energy effective continuum Hamiltonian, embodied by \cite{bitan-spin1, pal22b_berry, krish-spin1}
\begin{align} 
\label{eqham}
\mathcal{H}_\chi ( \boldsymbol k) & = 
\boldsymbol d ( \boldsymbol k) \cdot \boldsymbol  S  \,,
\quad 
\boldsymbol d( \boldsymbol k) =
\left \lbrace
\alpha_J \, k_\perp^J \cos(J\phi_k), \,
\alpha_J \, k_\perp^J \sin(J\phi_k), \,
 \chi \, v_z \, k_z \right \rbrace, 
 \quad J \in \lbrace 1, \, 2, \, 3  \rbrace , \nn
k_\perp & =\sqrt{k_x^2 + k_y^2}\,, \quad
\phi_k = \arctan({\frac{k_y}{k_x}}) ,
\quad \alpha_J=\frac{v_\perp}{k_0^{J-1}} \,.
\end{align}
Here, $\boldsymbol  S  = \lbrace S_x, \,S_y, \, S_z \rbrace$ represents the vector operator of the pseudospin-$1$ representation of the $SO(3)$ group, $\chi \in \lbrace 1, -1 \rbrace $ denotes the chirality of the node, and $v_z$ ($v_\perp$) is the Fermi velocity along the $z$-direction ($xy$-plane). The material-dependent parameter $k_0$, with the dimension of momentum, will have its value determined by the microscopic details of the material under consideration. We note that the conventional $J=1$ TSM is isotropic (forming an analogue of the isotropic WSMs), which is obtained by setting $ v_\perp = v_z$. For calculational purposes, we choose the representation where
\begin{align}
& S_x= \frac{1}{\sqrt{2}}
\begin{pmatrix}
	0 & 1  & 0  \\
	1 & 0 & 1  \\
	0 & 1 & 0  
\end{pmatrix} , \quad
S_y=\frac{1}{\sqrt{2}}
\begin{pmatrix}
	0 & -i  & 0  \\
	i & 0 & -i  \\
	0 & i & 0 
\end{pmatrix}, \quad
S_z =
\begin{pmatrix}
	1 & 0 & 0 \\
	0 & 0 & 0 \\
	0 & 0 &-1
\end{pmatrix}.
\end{align}
The energy eigenvalues of the Hamiltonian are given by
 \begin{align} 
\label{eigenvalues}
\varepsilon_ s  ({ \boldsymbol k})= 
 s \, \epsilon_{\boldsymbol k} \,, \quad
s \in \lbrace -1,0,1 \rbrace ,
\quad 
\epsilon_{\boldsymbol k}
= \sqrt{\alpha_J^2 \, k_\perp^{2J} + v_z^2 \, k_z^2}\,,
\end{align}
where the value $+1$ ($-1$) for $s$ represents the conduction (valence) band. The value zero represents a nondispersive flat-band. An orthogonal set of eigenvectors can be represented as follows:
\begin{align}
\label{eqev}
& \left\{-1, \, \frac {\chi  \,\sqrt {2} \,  k_z \,  v_z 
\, e^{i \, J \,\phi_k}}
 {\alpha_J \, k_\perp^{J} }, 
\,e^{2  \,i \, J \, \phi_k} \right\} \text{ for } s = 0 \text{ and } 
\nn &  \left\{
 1 + \frac {2 \, \chi \, k_z\, v_z 
\left (\chi  \, k_z\, v_z + s\, \epsilon_{\boldsymbol k}  \right)}
 {\alpha_J^2 \, \, k_\perp^{2 \, J}}, \,
 \frac {\sqrt {2}   \,e^{i  \,J \, \phi_k}
\left (\chi \,  k_z  \, v_z + s\, \epsilon_{\boldsymbol k}  \right)} 
{\alpha_J \, k_\perp^{J}}, \,
   \, e^{2  \,i \,  J\,\phi_k} \right\}
   \text{ for } s = \pm 1\,.
\end{align} 
Clearly, for the two dispersive bands, the energy varies (1) linearly along the $k_z$-direction, and (2) as $k_\perp^J$ when we confine to the $k_xk_y$-plane [cf. Fig.~\ref{figdis}(a)].
The group-velocity of the chiral quasiparticles, occupying the band with index $s$, is given by
\begin{align}
{\boldsymbol  v}_ s ( \boldsymbol k) 
\equiv \nabla_{\boldsymbol k} \varepsilon_ s   (\boldsymbol k)
=  \frac{s} {  \epsilon_{\boldsymbol k} }  
\left \lbrace J\, 
 \alpha_J^2 \,  k_\perp^{2J-2} \,  k_x , \, J \,  \alpha_J^2 \,  k_\perp^{2J-2} 
 \,  k_y  , \, v_z^2\,  k_z \right \rbrace .
\end{align}

\begin{figure*}[t]
\subfigure[]{\includegraphics[width=0.2 \linewidth]{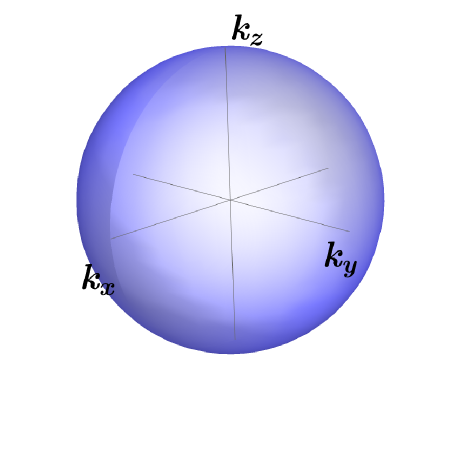} 
\includegraphics[width=0.2 \linewidth]{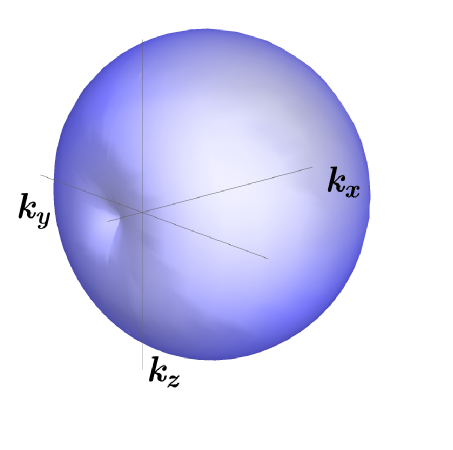}}
\hspace{ 1.5 cm}
\subfigure[]{\includegraphics[width=0.22 \linewidth]{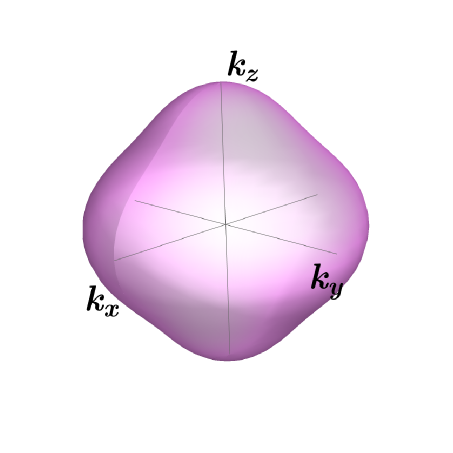}, \includegraphics[width=0.22 \linewidth]{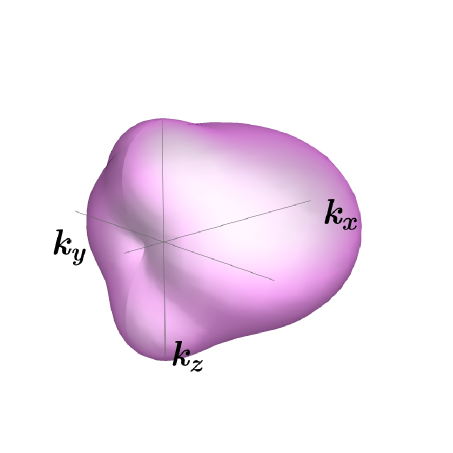}}
\caption{\label{figfs}Schematics of the Fermi surfaces for one node of a TSM with (a) J=1 and (b) J=2: In each subfigure, the two surfaces represent the case without and with the OMM-correction, respectively. Here, we have taken the applied magnetic field to be directed purely along the $x$-axis.
}
\end{figure*}

\subsection{Relevant topological quantities}

We discuss here the vectors given by the Berry curvature (BC) and the orbital magnetic moment (OMM), which will affect the linear response that we are set out to compute. For the band with index $s$, these are expressed by the generic formulae  of \cite{xiao_review,xiao07_valley}
\begin{align} 
\label{eqomm}
& {\boldsymbol \Omega}_\chi^s( \boldsymbol k)  = 
    i \, \langle  \nabla_{ \boldsymbol k}  \psi_\chi^s({ \boldsymbol k})| \, 
    \cross  \, | \nabla_{ \boldsymbol k}  \psi_\chi^s({ \boldsymbol k})\rangle 
\text{and } 
{\boldsymbol {m}}_\chi^s ( \boldsymbol k) = 
\frac{  -\,i \, e} {2 } \,
\langle  \boldsymbol \nabla_{ \boldsymbol k} \psi_\chi^s ({ \boldsymbol k})| \cross
\left [\,
\left \lbrace \mathcal{H}_\chi({ \boldsymbol k}) -\varepsilon_ s
({ \boldsymbol k}) 
\right \rbrace
| \boldsymbol \nabla_{ \boldsymbol k} \psi_\chi^s({ \boldsymbol k})\rangle \right ],
\end{align}
respectively. Here, $ \lbrace |  \psi_\chi^s({ \boldsymbol k}) \rangle  \rbrace $ is the set of normalized eigenvectors for the parent Hamiltonian.\footnote{For example, the set shown in Eq.~\eqref{eqev} can be used after normalization.} On evaluating the expressions in Eq.~\eqref{eqomm} using  $\mathcal{H}_\chi ( \boldsymbol k)$, we get
\begin{align}
& \boldsymbol \Omega_\chi^s ({ \boldsymbol k})= 
 \frac{ -\, \chi \, s \,
J \,v_z \, \alpha_J^2 \, k_\perp^{2J-2} }
{\epsilon^3_{\boldsymbol k}
} \left
\lbrace k_x, \, k_y, \, J\, k_z \right \rbrace , \nn
&   {\boldsymbol {m}}_\chi^s ({ \boldsymbol k}) 
=   \frac{ -\, \chi \, e\, J \,v_z \,\mathcal{G}_ s\,  \alpha_J^2 \, k_\perp^{2\,J-2} }
{2 \, \epsilon^2_{\boldsymbol k}} 
 \left \lbrace k_x, \, k_y, \, J \, k_z \right \rbrace,
 \text{ where } 
 \mathcal{G}_ s =\begin{cases}
 1 & \text{ for } s=\pm 1 \\
 2 & \text{ for } s=0
 \end{cases} \, .
\end{align}
For the flat-band, although the BC is identically zero, 
the OMM is nonzero and turns out to be twice the OMM of either of the dispersive bands.
It is easy to verify from the $\boldsymbol \Omega_\chi^s ({ \boldsymbol k})$-expressions that the node has a net Chern number of  $2\, \chi \,J$. 

Due to the zero dispersion and zero BC for the flat-band, it leads to zero conductivity when OMM is ignored. Henceforth, we will then neglect the flat-band, as far as our calculations of linear response are concerned. We note that, for the dispersive bands, while the BC changes sign with $s$, the OMM does not. Therefore, for uncluttering of notations, we will henceforth remove the superscript ``$s$'' from ${\boldsymbol {m}}_\chi^s({ \boldsymbol k})$.

The first and foremost effect of BC is that it modifies the phase-space volume element via a factor of $
\left [{\mathcal D}_\chi^s  (\boldsymbol k)\right]^{-1}$, where
\begin{align}
{\mathcal D}_\chi^s  (\boldsymbol k) = \left [1 
+ e \,  \left \lbrace 
{\boldsymbol B} \cdot \boldsymbol{\Omega }_\chi^s  (\boldsymbol k)
\right \rbrace  \right ]^{-1}.
\end{align}
A nonzero OMM causes a Zeeman-like correction to be added to the dispersion \cite{xiao_review}, leading to the effective dispersion of
\begin{align}
\label{eqmodi}
& \xi_\chi^s (\boldsymbol k) 
= \varepsilon_ s  (\boldsymbol k) + \varepsilon_\chi^{ (m) }  (\boldsymbol k) \, ,
\quad 
\varepsilon_\chi^{(m)}   (\boldsymbol k) 
= - \,{\boldsymbol B} \cdot \boldsymbol{m }_\chi  (\boldsymbol k) \,.
\end{align}
This, in turn, modifies the group-velocity as
\begin{align}
{\boldsymbol   w}_\chi^s ({\boldsymbol k} ) \equiv 
 \nabla_{{\boldsymbol k}}   \xi_\chi^s ({\boldsymbol k})
 = {\boldsymbol   v}_ s ({\boldsymbol k} ) + {\boldsymbol   u}^{(m)}_\chi ({\boldsymbol k} ) \,,
\quad {\boldsymbol  u}^{(m)}_\chi ({\boldsymbol k} )
= \nabla_{{\boldsymbol k}} \varepsilon_\chi^{(m)}   (\boldsymbol k) \,.
\end{align}
The modified effective Fermi surface, on including the OMM-correction, is illustrated schematically in Fig.~\ref{figfs}.

The effects of OMM will be show up in the equilibrium Fermi-Dirac distribution,
\begin{align}
\label{eqdist}
	f_0 \big (\xi^\chi_s (\boldsymbol k) , \mu, T (\boldsymbol r) \big )
= \frac{1}
{ 1 + \exp [ \, 
\left(  \xi^\chi_{s} (\boldsymbol k)-\mu \right) \,\beta  (\boldsymbol r)\,  ]}\,,
\end{align}
where $\beta (\boldsymbol{r}) =  1/T (\boldsymbol{r})$.
While using $f_0$ in various equations, we will be suppressing its $\mu$- and $ T $-dependence for uncluttering of notations.

\subsection{Expansion in $B$}

In order to obtain closed-form analytical expressions, we will expand the $B $-dependent terms upto a given order in $B$, assuming it has a small magnitude. This is anyway an essential condition in order to neglect the quantization of the dispersion into discrete Landau levels, and for applying the semiclassical Boltzmann formalism using the effective dispersion of Eq.~\eqref{eqmodi}. More specifically, we must have a small cyclotron frequency, $\omega_c=e\,B/m^*$ (where $m^* $ is the effective mass with the magnitude $\sim 0.11 \, m_e$ \cite{params2}, with $m_e$ denoting the electron mass). The regime of validity of our calculational framework holds when $  \omega_c \ll \mu$, where $\mu$ is the energy at which the chemical potential cuts a dispersing band.

The weak-magnetic-field limit implies that 
\begin{align}
 e \, |{\boldsymbol B} \cdot \boldsymbol{\Omega }_\chi^s  | \ll 1 
 \text{ and }
\left| \varepsilon_\chi^{(m)}   (\boldsymbol k) \right |
 \ll  |\varepsilon_s | . 
 \label{eqcond}
\end{align}
In what follows, we will calculate all the terms upto $\order{B^2}$. This implies that
we use the following expansions:
\begin{align}
{\mathcal D}_\chi^s &=
 1 - e \,  \left( {\boldsymbol B}  \cdot \boldsymbol{\Omega }_\chi^s  \right) 
+   e^2  \,  \left( {\boldsymbol B} \cdot \boldsymbol{\Omega }_\chi^s  \right)^2  
+  \order{ B^3} \,,\quad
 f_{0} ( \xi^s_\chi  ) = f_{0} (\varepsilon_s ) 
+  \varepsilon_\chi ^{ (m) } \,  
f^\prime_{0}(\varepsilon_s ) 
+ \frac{1}{2} \left(  \varepsilon_\chi^{(m)} \right)^2  
  f^{\prime \prime}_{0}(\varepsilon_s )  +  \order{ B^3} \,.
 \label{eqexp}
\end{align}
Here, the ``prime'' superscript denotes partial-differentiation with respect to the variable shown explicitly within the brackets [e.g., $ f_0^\prime (\varepsilon) \equiv \partial_\varepsilon f_0 (\varepsilon)$].

\subsection{Linear-response coefficients}
\label{seclinear}

Derived through the semiclassical Boltzmann equations, we will investigate the transport properties in the linear-response regime, applicable for small values of the probe fields of $\boldsymbol E$ and $ \nabla_{\boldsymbol r} T$. Our planar-Hall and planar-thermal-Hall configurations are shown in Fig.~\ref{figdis}(b).
In the following sections, we will compute the resulting three linear-response coefficients, $\sigma_\chi (s) $, $\alpha_\chi (s) $, and $\ell_\chi (s) $, whose physical significance can be understood from the discussions below Eq.~\eqref{eqcur1}. We will consider a positive chemical potential $ \mu $ being applied to the node, such that the Fermi level cuts the positive-energy band.
Since the steps to obtain the forms of linear-response coefficients have been extensively discussed in Refs.~\cite{ips_rahul_ph_strain, ips-kush-review, ips-rsw-ph}, we do not review it here. We just use the final answers, which are explained below.

The linear-response coefficients relate the average electric and thermal current densities, ${\boldsymbol J}_\chi^s $ and ${\boldsymbol J}^{s, {\rm th}}_\chi$, contributed by fermionic quasiparticles associated with the band $s$ and node with chirality $\chi$, to the driving electric potential gradient and temperature gradient. This relation is expressed in a compact form as \cite{mermin}
\begin{align}
\label{eqcur1}
\begin{bmatrix}
\left( J_\chi^s\right)_i \vspace{0.2 cm} \\
\left( {J}_\chi^{s, {\rm th}}\right)_i 
\end{bmatrix} & = \sum \limits_j
\begin{bmatrix}
 \left( \sigma_\chi \right)_{ij} (s) &  \left( \alpha_\chi \right)_{ij}  (s) 
\vspace{0.2 cm}  \\
T  \left( \alpha_\chi \right)_{ij}  (s)  &  \left( \ell_\chi \right)_{ij}  (s) 
\end{bmatrix}
\begin{bmatrix}
E_j
\vspace{0.2 cm}  \\
- \, { \partial_{j} T } 
\end{bmatrix} ,
\end{align}
where $ \lbrace i, j \rbrace  \in \lbrace x,\, y, \, z \rbrace $ indicates the Cartesian components of the current vectors and the response tensors in 3d. While $\sigma_\chi (s) $ and $\alpha_\chi (s) $ represent the magnetoelectric conductivity and the magnetothermoelectric conductivity, respectively, $\ell_\chi (s)  $ represents the tensor relating the thermal current density to the temperature gradient, at a vanishing electric field.
Since it is used to compute the magnetothermal coefficient~\cite{mermin, ips-kush-review, ips-hermann-review, ips-rsw-ph}, $\kappa_\chi (s)$, we will loosely refer to $\ell_\chi (s) $ itself as the magnetothermal coefficient.

In this subsection, we discuss the scenario when only the intranode scatterings play a dominant role, implemented through using a phenomenological momentum-independent relaxation time ($\tau$).
From the final expressions obtained from the \textit{linearized} Boltzmann equations \cite{ips-ruiz, ips-rsw-ph}, we divide up the electric conductivity into three parts as
\begin{align}
\label{eqsigmatot}
& \sigma_\chi (s)
=\bar \sigma_\chi (s) + \sigma_\chi^{\rm AH} (s)
+ \sigma_\chi^{\rm LF} (s) \,.
\end{align}
The contents and significance of these three parts are described below:
\begin{enumerate}

\item The first part arises from the current density of
$
\bar {\boldsymbol J}_\chi^s = 
- \, e^2 \, \tau 
\int \frac{ d^3 \boldsymbol{k}} {(2 \, \pi)^3} \, 
\mathcal{D}_\chi^s  \left( {\boldsymbol w}_\chi^s 
+ {\boldsymbol W}_\chi^s \right )
\left (  {\boldsymbol w}_\chi^s  +
 {\boldsymbol W}_\chi^s \right ) \cdot \boldsymbol{E} \;
 f^\prime_0 (\xi_\chi^s) \,,
$
and takes the form of
\begin{align}
\label{eqbarsig}
    \left(\bar \sigma_\chi \right)_{i j} (s)
&= - \,e^2 \, \tau  
\int \frac{ d^3 \boldsymbol k}{(2\, \pi)^3 } \, \mathcal{D}_\chi^s 
\left[  (w_\chi^s)_i \, + (W_\chi^s)_i \right ]
\left [ (w_\chi^s)_j \, + (W_\chi^s)_j \right] \, f^\prime_0 (\xi_\chi^s)  \,,\quad
\boldsymbol{W}_\chi^s 
= e \left  ( {\boldsymbol w}_\chi^s \cdot 
  \boldsymbol {\Omega}_\chi^s \right  ) \boldsymbol{B}\,.
\end{align}
It comprises only even powers of $ B $ and has only nonzero in-plane components (i.e., the out-of-plane components vanish).

\item 
The second part comes from the electric current density of
$
{\boldsymbol J}_\chi^{s,\rm AH}  =   -\, e^2   \int
\frac{ d^3 \boldsymbol k}{(2\, \pi)^3 } \,
\left   ({\boldsymbol E}  \cross   \boldsymbol \Omega^s_\chi   \right)
  f_0  (\xi_\chi^s) \,,
$ 
which gives rise to the ``intrinsic anomalous-Hall effect'' \cite{haldane04_berry, goswami13_axionic, burkov14_anomolous}. Hence,
\begin{align}
\left( \sigma_\chi^{\rm AH}\right)_{ij} (s)
&= -\, e^2  \,\epsilon_{ijl} 
\int \frac{ d^3 \boldsymbol k}{(2\, \pi)^3 } \, \left(\Omega_\chi^s \right)^l 
\,  f_0  (\xi_\chi^s) \,,\nn
\end{align}
with its longitudinal component evaluating to zero (due to the presence of the Levi-Civita symbol). This part is completely independent of the relaxation time $\tau $. If OMM is set to zero, $ \sigma_\chi^{{\rm AH}} (s)$ becomes independent of $
\boldsymbol  B$ and, thus, 
vanishes identically. We also note that, for our set-up with the applied fields and temperature gradient confined to the $xy$-plane, the in-plane transverse component [i.e., $\left( \sigma_\chi^{\rm AH}\right)_{yx} (s)$]  also vanishes, with only the out-of-plane transverse component surviving.

\item 
The third part is the so-called Lorentz-force part, and it arises from the current density of
\cite{ips-rsw-ph, ips_tilted_dirac}
 \begin{align} 
 \label{eqcurlf} 
& {\boldsymbol J}^{s, \rm LF}_\chi = 
-\,e^2 \,  \tau \int \frac{d^3 \boldsymbol{k}} {(2 \, \pi)^3} 
\left(  {{\boldsymbol w}}_\chi^s 
+   {{\boldsymbol W}}_\chi^s \right )
\, f_{0}^\prime (\xi_\chi^s)\, \mathcal{Y}_\chi^s ,
  \text{ where } \check{L} 
  = ({\boldsymbol w}_\chi^s \cross \boldsymbol{B}) \cdot \nabla_{\boldsymbol{k}}
\nn &
\text{and } \mathcal{Y}_\chi^s =  
\sum_{n = 1}^{\infty}
\left (e \, \tau \, \mathcal{D}_\chi^s \right )^n \check{L}^n 
\left [    \mathcal{D}_\chi^s \left
\lbrace {{\boldsymbol w}}_\chi^s 
+  {{\boldsymbol W}}_\chi^s  \right 
\rbrace \cdot \boldsymbol{E}   \right ].  
\end{align} 
This part arises from the action of the Lorentz-force operator, $\check{L} $, and the solution is obtained
by expanding the summation series upto a certain value of $n$.
The nomenclature reflects the fact that it includes the classical Hall effect due to the Lorentz force. The corresponding components of the electric conductivity is
 \begin{align} 
\label{eqsiglf} 
\left(\sigma_\chi^{\rm LF} \right)_{i j} (s) & = 
-\,e^2 \,  \tau \int \frac{d^3 \boldsymbol{k}} {(2 \, \pi)^3} 
\, \left[  (w_\chi^s)_i  + (W_\chi^s)_i \right ]
\, f_{0}^\prime (\xi_\chi^s) \,
\frac{\partial \mathcal{Y}_\chi^s } {\partial E_j}\,.
\end{align}

\end{enumerate}

Regarding the magnetothermoelectric conductivity and the magnetothermal coefficient, we will only explicitly show the expressions for the in-plane components which arise from the non-anomalous-Hall and non-Lorentz-force parts (of the corresponding current densities). The relevant quantities are obtained from evaluating~\cite{ips-ruiz, ips-rsw-ph}
\begin{align} 
\label{eqbaralpha}
	\left (\bar{\alpha}_\chi \right )_{ij} (s) &= 
e \, \tau  \int \frac{d^3 \boldsymbol{k}} {(2 \, \pi)^3} \,\mathcal{D}_\chi^s  
\left [ (w_\chi^s)_i +  (W_\chi^s)_i \right ]
\left [ (w_\chi^s)_j +  (W_\chi^s)_j \right ]
\frac{(\xi_\chi^s - \mu  )}{T}
	   \, f^\prime_0 (\xi_\chi^s) 
\end{align}
and
\begin{align}
\label{eqbarell}
\left ( {\bar \ell}_\chi \right )_{ij} (s)
=  - \,\tau   \int \frac{d^3 \boldsymbol{k} }  {(2 \, \pi)^3} \,
\mathcal{D}_\chi^s \, \frac{  (\xi_\chi^s - \mu )^2}{T} 
\left[  (w_\chi^s)_i +  (W_\chi^s)_i \right ] 
\left[ (w_\chi^s)_j +  (W_\chi^s)_j \right ]
f^\prime_0 (\xi_\chi^s)\,,
\end{align}
respectively.

In the following sections, we will demonstrate the explicit expressions for the three independent linear-response coefficients, viz. $  \sigma_\chi$, $\alpha_\chi$, and $  \ell_\chi$. We assume a positive chemical potential $ \mu $ being applied to the node in question, such that the Fermi level cuts the conduction band, which contributes to transport. Therefore, $s$ is set equal to one and, henceforth, we will suppress the $s$-dependence.

\section{Magnetoelectric conductivity} 
\label{secsigma}

The generic expressions for the in-plane components of the magnetoelectric conductivity, obtained by expanding Eq.~\eqref{eqbarsig} (in powers of $B$) upto $\order{B^2}$, can be found in Ref.~\cite{ips-ruiz}. Here, we only show the final results applicable for TSMs. In fact, since the in-plane components contain only even powers of $B$, the expressions shown below are correct upto $\order{B^3}$. Notationwise, the superscript ``Drude'' refers to the 
 $B$-independent parts. Furthermore, the superscripts of ``BC'' and ``$m$'' are used to indicate whether the OMM contributions have been set to zero or not.
We also compute and demonstrate the expressions for the parts arising from the anomalous-Hall and Lorentz-force parts, which contain only odd powers of $B$, and are correct upto $\order{B^3}$. Hence, overall, all our expressions are correct upto $\order{B^3}$.

\subsection{Longitudinal component of $\bar \sigma_\chi$}

\begin{figure*}[t]
\includegraphics[width=0.95 \linewidth]{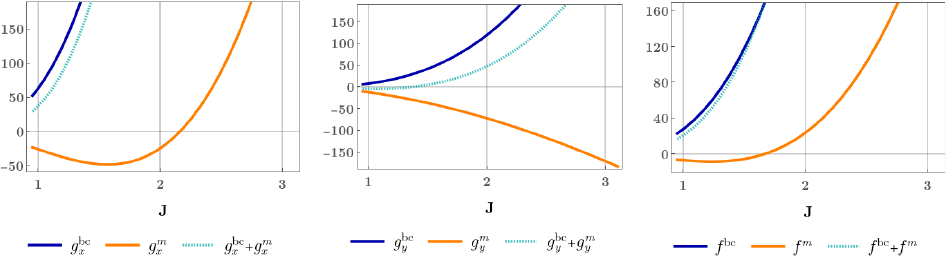}
\caption{\label{figcom}Comparison of the values of the functions defined in Eqs.~\eqref{eqg} and \eqref{eqf}  for $J=1, \,2, \,3$.}
\end{figure*}

The part-by-part expressions for the longitudinal component of the electric conductivity are given by
\begin{align}
\label{eqbarsxx}
&\left( \sigma^{\text{Drude}}_\chi  \right)_{xx} =\frac{ e^2 \,  \tau  \,  J   }
{ 6 \,  \pi^2   \,  v_z    }  \,  \Upsilon_{2} (\mu, T ) \,,
\left( \sigma^{\text{BC}}_\chi  \right)_{xx}=
\frac{e^4 \,  \tau \,  v_{z } \,  \alpha_{J}^{ \frac{2} {J} }   
\, \Upsilon_{ - \frac{2} {J} } ( \mu, T )}
{ 128 \,  \pi ^{\frac{3}{2}} } \,  
\frac{  \Gamma  \big( \frac{2 \, J - 1} {J} \big) }
{  \Gamma \big (  \frac{9\, J - 2} {2\, J}  \big)} 
 \left [   g^{bc}_x (J)\,B_x^2 
 +   g^{bc}_y (J) \,B_y^2 \right ],
\nn &
(\sigma_\chi^{m})_{xx}= 
\frac{e^4 \,  \tau \,  v_{z } \,  \alpha_{J}^{ \frac{2} {J} }   
\, \Upsilon_{ - \frac{2} {J} } ( \mu, T )
}
{ 128 \,  \pi ^{\frac{3}{2}} } \,  
\frac{  \Gamma  \big( \frac{2 \, J - 1} {J} \big) }
{  \Gamma \big ( \frac{9\, J - 2} {2\, J} \big)}
 \left [   g^{m}_x (J)\,B_x^2 
 +   g^{m}_y (J) \,B_y^2 \right ] ,
\end{align}
where
\begin{align}
\label{eqg}
& g^{bc}_x   (J) = 4\,J \left ( 32 \,  J^2 -19  \, J + 3  \right ) , \quad
g^{bc}_y  (J) = 4\, J \left (3\, J- 1 \right) \left  (2\,J-1 \right)  , \nn &
g^{m}_x (J)=  \frac{59 \, J^4 - 175\, J^3 + 115 \, J^{2} -27 \, J +2 } { J} \,,\quad
 g^{m}_y  (J) =  \frac{J^4 - 25\,J^3 +7 \,J^2  + 7\, J - 2} { J}  \,,
\end{align}
and
\begin{align}
\label{eqsomfac}
 \Upsilon_{n} (\mu, T ) 
=  \mu^{n} \,  \left[ 1 + \frac{\pi^2 \,  n \,  (n-1)}
{6 } \, \left( {T}/{\mu}\right)^2 
+ \order{ \left( {T}/{\mu}\right)^3 } \right]
\text{[arising from the Sommerfeld expansion shown in Eq.~\eqref{eqsom}]} .
\end{align}
Their behaviour is depicted in Fig.~\ref{figcom}. In particular, we find that
\begin{align*}
& \lbrace g^{bc}_x(1) , \, g^{bc}_x(2), \, g^{bc}_x(3)  \rbrace 
= \lbrace 64, 744, 2808
  \rbrace\,, \quad
\lbrace g^{m}_x(1) , \, g^{m}_x(2), \, g^{m}_x(3)  \rbrace 
= \lbrace -26, -24, 1010/3
  \rbrace \,, \nn &
\lbrace g^{bc}_y(1) , \, g^{bc}_y(2), \, g^{bc}_y(3)  \rbrace 
= \lbrace 8, 120, 480
 \rbrace  \,, \quad
\lbrace g^{m}_y(1) , \, g^{m}_y(2), \, g^{m}_y(3)  \rbrace 
= \lbrace -12, -72, -512/3
 \rbrace\,.
\end{align*}
This leads to following conclusions:
\\(1) $g_x^{bc}  (J)$ and $g_y^{bc} (J)$ are positive for all $J$-values.
\\(2) $g_x^m (J)$ is negative for $J=1 $ and $J=2$, and positive for $J=3$.
\\(3) $g_y^m (J)$ is negative for all $J$-values.
\\Therefore, for $J=1 $ and $J=2$, the OMM acts in opposition to the BC-only term for the $B_x^2$-dependent part, thus reducing the intensity of the overall response. On the other hand, for $J=3$, the OMM-part adds up to the BC-only term, thus increasing the overall response for the $B_x^2$-dependent part. However, for all values of $J$, the addition of the OMM does not change the sign of the overall response.
Next, considering the $B_y^2$-dependent part, we find that the BC-only and the OMM parts have opposite signs for all values of $J$. While the OMM-part manages to flip the sign of the overall response for $J=1$, its magnitude is too low to do so for $J=2 $ and $J=3$. All these values can be compared with the results for WSMs and mWSMs, shown in Ref.~\cite{ips-ruiz}.

\subsection{In-plane transverse component of $\bar \sigma_\chi$}

The part-by-part expressions for the in-plane transverse component are given by
\begin{align}
\label{eqbarsyx}
& \left( \sigma^{\text{Drude}}_\chi \right)_{yx} =0 \,, \quad
\left( \sigma^{\text{BC}}_\chi \right)_{yx}  =
  \frac{e^4 \,  \tau \,  v_z \,    \alpha_{J}^{ \frac{2}{J} } 
 \,   \Upsilon_{ - \frac{2} {J} } (\mu, T )
 }
{ 64 \,  \pi ^{\frac{3}{2}}  }    \,   \frac{\Gamma \big (  \frac{2 \, J - 1} {J}   \big)}
{\Gamma \big (  \frac{9\, J - 2} {2\, J} \big)}
  \, f^{bc} (J) \, B_x \, B_y\,,
\nn &
(\sigma_\chi^{m})_{yx} = 
  \frac{e^4  \,  \tau \, v_z
\, \alpha_{J}^{ \frac{2} {J} }  \,  \Upsilon_{ - \frac{2} {J} } (\mu, T )    }
{ 64 \,  \pi ^{\frac{3}{2}} } \, 
  \frac{  
  \Gamma \big (  \frac{2 \, J - 1} {J}   \big)}
  {  \Gamma \big (  \frac{9\, J - 2} {2\, J} \big)} 
 \, f^{m} (J) \, B_x \, B_y\, ,
\end{align}
where
\begin{align}
\label{eqf}
& f^{bc} (J) = 4\,J \left ( 13\, J^{2} - 7 \, J +1 \right ) , \quad
f^m(J) =29 \, J^3- 75 \,J^2 + 54\, J -17 + \frac{2}{J}\, .
\end{align}
Their behaviour is depicted in Fig.~\ref{figcom}. In particular, we find that
\begin{align*}
& \lbrace f^{bc}(1) , \, f^{bc}(2), \, f^{bc}(3)  \rbrace 
= \lbrace 28, 312, 1164  \rbrace\,, \quad
\lbrace f^{m}(1) , \, f^{m}(2), \, f^{m}(3)  \rbrace 
= \lbrace -7, 24, 761/3  \rbrace\,.
\end{align*}
This leads to following conclusions:
For $J=1 $, the BC-only and the OMM parts have opposite signs and, thus, the OMM-part reduces the magnitude of the overall response. For each of $J=2$ and $J=3$, the OMM-part adds up to the BC-only term, thus reinforcing each other. However, for all values of $J$, the addition of the OMM does not change the sign of the overall response.
All these values can be compared with the results for WSMs and mWSMs, shown in Ref.~\cite{ips-ruiz}.

\subsection{Out-of-plane component due to the anomalous-Hall effect}
\label{secah_ah}

For the intrinsic anomalous-Hall part, we actually need to expand $f_0 (\xi^s_\chi)$ upto order $B^3$, because it comprises only odd powers of $B$ and we, ultimately, want to include the results for all components correct upto order $B^3$ (see the appendix of Ref.~\cite{ips-rsw-ph} for more details). Therefore, using
$$ f_{0} ( \xi^s_\chi  ) = f_{0} (\varepsilon_s) 
+  \varepsilon_\chi ^{ (m) } \,  
f^\prime_{0}(\varepsilon_s ) 
+ \frac{1}{2} \left(  \varepsilon_\chi^{(m)} \right)^2  
  f^{\prime \prime}_{0}(\varepsilon_s )
  + \frac{1}{6} \left(  \varepsilon_\chi^{(m)} \right)^3  
 f^{\prime \prime \prime}_{0}(\varepsilon_s)  +  \order{ B^4},$$
the application of the Sommerfeld expansion [cf. Eq.~\eqref{eqsom}] yields
\begin{align}
\label{eqah}
(\sigma_\chi^{ {\text{AH}}})_{zx}  & =
\frac{-\,  e^3 \, v_z \, J \,  B_y }
	 {2 \, \pi^2}
\left [ \frac{\Upsilon_{-1}(\mu ,T)}{6} 
+ 
\frac{  
B^2 \,  e^2 \, v_z^2 \,  \alpha_{J}^{ \frac{2} {J} } \,\sqrt{\pi }\;
\Gamma  \big(  \frac{4\, J-1} {J}\big) \, h_J } 
{128 \, \Gamma  \big( \frac{9 \, J -2} {2 \, J}\big)}
 \, \Upsilon_{ \frac{-3\, J - 2} {J}}(\mu ,T) \right ],  
\quad h_J  =    J \,  \left( J+1 \right) \left( J+2 \right).
\end{align}
It does not have any BC-only contribution.
Let us compare the above expression with the one for a single node of the WSM/mWSM variety. On carrying out explicit computations using the Hamiltonian shown in Ref.~\cite{ips-ruiz}, we find it to be
\begin{align}
\label{eqahmweyl}
(\sigma_\chi^{ {\text{AH}}})_{zx} \Big \vert_{\text{mWSM}} & =  
\frac{ -\,e^3 \, v_z \, J \,  B_y }
	 {2 \, \pi^2}
\left [ \frac{\Upsilon_{-1}(\mu ,T)}{12} 
+ 
\frac{  
B^2 \,  e^2 \, v_z^2 \,  \alpha_J^{ \frac{2} {J} } \,\sqrt{\pi }\,
\Gamma  \big(  \frac{4\, J-1} {J}\big) \, h_J } 
{256 \,  \Gamma  \big( \frac{9 \, J -2} {2 \, J}\big)}
 \, \Upsilon_{ \frac{-3\, J - 2} {J}}(\mu ,T) \right ] .
\end{align}
The analogous contribution for a single RSW node can be found in Ref.~\cite{ips-rsw-ph}.

\subsection{Part of the conductivity arising from the Lorentz-force operator}
\label{seclf}

We divide up the part of the conductivity, $\sigma_{ \chi }^{ {\text{LF}}} $, arising from the Lorentz-force operator [cf. Eq.~\eqref{eqsiglf}] into three subparts, which represent the contributions arising (1) independent of the topological properties like BC and OMM (this one includes the part giving rise to the conventional Hall effect); (2) purely from the BC (i.e., when OMM is neglected), and (3) when OMM is included: 
\begin{align}
\label{eqlf}
& \sigma_{ \chi }^{ {\text{LF}}} =
\sigma_\chi^{\text{LF}, \text{H}} +
\sigma_\chi^{\text{LF}, \text{BC}} +
\sigma_\chi^{\text{LF},m} \,.
\end{align}
Appendix~\ref{appLF} contains more details regarding the derivations of the generic expressions, expanded upto order $B^3$. An important point to notice is that the Lorentz-force operator gives rise to in-plane components, in addition to the out-of-plane ones, which are evident only when we consider terms arising from $n\geq 2$ from the summation series shown in Eq.~\eqref{eqcurlf}. Let us summarize our results below:
 \begin{enumerate}
\item{$n=1$:}
\\Using the generic expression shown in Appendix~\ref{appn1}, we get
nonzero values only for the out-of-plane parts. They take the forms of
\begin{align}
& \left (\sigma_\chi^{\text{LF}, \text{H}} \right )_{zx} =
\frac{ -\,e^3 \, v_z \, J \, \tau^2 \, B_y }
{6 \, \pi^2} \, \Upsilon_{1}(\mu ,T) \,, \quad
\left  (\sigma_\chi^{\text{LF}, \text{BC}} \right )_{zx} =
  \frac{ -\,3\,  J^3\,
  e^5 \, v_z^3 \, \tau^2 \, \alpha_{J}^{ \frac{2} {J}} \, B_y\, B^2 \,}
	 {16 \, \pi^{\frac{3}{2}}} \,
\frac{ \Gamma  \big( \frac{3 \, J - 1} {J} \big)}
{\,  \Gamma  \big(\frac{ 9 \, J -2 } {2 \, J}\big)
}  
\, \Upsilon_{ -\frac{J+2}{J}}(\mu ,T) \,, \nn
& \left (\sigma_\chi^{\text{LF},m} \right )_{zx} =  
\frac{ e^5 \, v_z^3 \, \tau^2 \, 
\alpha_{J}^{ \frac{2} {J}} \, B_y\, B^2 \,}
	 {128 \, \pi^{\frac{3}{2}}} \,
\frac{ \Gamma  \big( \frac{3 \, J - 1} {J} \big)}
{\,  \Gamma  \big(\frac{ 9 \, J -2 } {2 \, J}\big)
}\,
\mathcal{L}_J^{ m }\,
\Upsilon_{ -\frac{J+2}{J}}(\mu ,T) \,,
\text{ where }
\mathcal{L}^m_J = 10\, J^3 + 21\, J^2+J-2 \,.
\end{align}

\item{$n=2$:}
\\For the generic expression shown in Appendix~\ref{appn2}, when we evaluate the integrals, we find that
only in-plane components appear as the nonzero parts. They take the forms of
\begin{align}
\label{eqlfxx}
& \left(\sigma_\chi^{\text{LF}, \text{BC}} \right )_{xx} = 
\left  (\sigma_\chi^{\text{LF}, m} \right )_{xx}=0\,,
\quad \left (\sigma_\chi^{\text{LF}, \text{H}} \right )_{xx}
=
\frac{- \,e^4 \,  \tau^3 \,  v_{z } \,  
\alpha_{J}^{ \frac{2} {J} }   
\, \Upsilon_{\frac{2\, J -2} {J} } ( \mu, T )
}
{16 \,  \pi^{\frac{3}{2}} } \,  
\frac{  
\Gamma  \Big( \frac{2 \, J - 1} {J} \Big) }
{  \Gamma \Big (  \frac{7 \, J -2 }{2\, J}   \Big)} 
 \left [   n_x (J) \,B_x^2  +   n_y (J) \,B_y^2 \right ],\nn
&\text{where }
  n_x   (J) = J \left ( J  -1 \right )^2 , \quad
n_y  (J) = 3\, J^3 -2\, J^2+ J \,;
\end{align}
and
\begin{align}
\label{eqlfyx}
&  \left( \sigma^{\text{LF}, \text{BC}}_\chi \right)_{yx} 
= \left (\sigma_\chi^{\text{LF}, m} \right )_{yx} =0 \,,\quad
 \left( \sigma^{\text{LF},\text{H}}_\chi \right)_{yx} 
=
\frac{B_x\, B_y\, e^4\, J^3\, \tau ^3 \, v_z \, \alpha_J^{\frac{2}{J} }  
\, \Upsilon_{\frac{2\, J -2} {J} }} {8 \, \pi^\frac{3}{2} } 
\frac{  \Gamma  \Big( \frac{2 \, J - 1} {J} \Big) }
{  \Gamma \Big (  \frac{7 \, J -2 }{2\, J}   \Big)} \,.
\end{align}
We observe that the in-plane components are generated exclusively from the BC- and OMM-independent contributions coming from Eq.~\eqref{eqb12}.

\item{$n=3$:}
\\On evaluating the integrals using the generic expression shown in Appendix~\ref{appn3}, we conclude that
only the $zx$-component survives. It is given by
\begin{align}
\label{eqLFtot}
& \left( \sigma_\chi^{\text{LF}, \text{H}} \right )_{zx}=  
\frac{ e^5 \, v_z^3 \, \tau^4 \, \alpha_{J}^{ \frac{2} {J}} \, B_y\, B^2 \,}
	 {16 \, \pi^{\frac{3}{2}}} \,
\frac{ \Gamma  \big( \frac{2 \, J - 1} {J} \big)}
{ \Gamma  \big( \frac{7\, J -2} {2\, J} \big)
} 
\left (3 \, J^3 -2 \,J^2 +J \right)
\Upsilon_{ \frac{ J -2} {J}}(\mu ,T) \, .
\end{align} 
Similar to the terms under $n=2$, the $n=3$ term is generated exclusively from the BC- and OMM-independent contributions. In this case, they arise from Eq.~\eqref{eqb15}.

\end{enumerate}
Our results show that only currents proportional to odd powers of $B$ and even powers of $\tau$ survive for the out-of plane component. On the other hand, the in-plane components contain only even powers of $B$ and odd powers of $\tau$. Gathering all the contributions shown above, we get the total for the out-of-plane part as
\begin{align}
&   \left (\sigma_{ \chi }^{ {\text{LF}}} \right )_{zx}   =
\frac{ -\,
e^3 \, v_z \, J \, \tau^2 \, B_y }
	 {2 \, \pi^2} \,
\Bigg [ \frac{\Upsilon_{1}(\mu ,T) }{3} 
+ 
\frac{ e^2 \,  v_z^2 \, B^2 \, \alpha_{J}^{ \frac{2} {J} } \,\sqrt \pi  }
{8}
\, \Bigg \lbrace 
\frac{ \Gamma  \big( \frac{3 \, J - 1} {J} \big)}
{8\,\Gamma  \big( \frac{9 \, J - 2} {2\,J} \big)}\, 
 \mathcal{L}_{J,1}
\Upsilon_{ - \frac{ J+2}{J}}(\mu ,T) 
 - \frac{\tau^2\, \Gamma  \big( \frac{2\, J - 1} {J} \big)}
 { \Gamma  \big( \frac{7 \, J - 2} {2\,J} \big)}\,  \mathcal{L}_{J,2}
 \, \Upsilon_{\frac{ J -2}{J}}(\mu ,T)
 \Bigg \rbrace \Bigg ]\, ,\nn
& \text{where }
  \mathcal{L}_{J,1} =14 \, J^2 - 21\, J   
-1+\frac{2}{J} \text{ and }
 \mathcal{L}_{J,2} = 3\, J ^2-2\, J+1 \,.
\end{align}

\subsubsection{Comparison with WSMs/mWSMs}

Using the Hamiltonian shown in Ref.~\cite{ips-ruiz}, we now compare TSMs' behaviour with the results we derive for the WSMs/mWSMs.
For a conduction band, the individual terms arising from $n=1, 2, 3$ are shown below:
 \begin{enumerate}
\item{$n=1$:}
\begin{align}
& \left (\sigma_\chi^{\text{LF}, \text{H}} \right )_{zx} =
\frac{ -\, e^3 \, v_z \, J \, \tau^2 \, B_y }
{6 \, \pi^2} \, \Upsilon_{1}(\mu ,T) \,, \quad
\left  (\sigma_\chi^{\text{LF}, \text{BC}} \right )_{zx} 
=
 \frac{ -\, 3 \,J^3
 \, e^5 \, v_z^3 \, \tau^2 \, \alpha_{J}^{ \frac{2} {J}} \, B_y\, B^2 \,}
	 { 64\, \pi^{\frac{3}{2}}} \,
\frac{ \Gamma  \big( \frac{3 \, J - 1} {J} \big)}
{\,  \Gamma  \big(\frac{ 9 \, J -2 } {2 \, J}\big)
}  
\, \Upsilon_{ -\frac{J+2}{J}}(\mu ,T) \,, \nn
& \left (\sigma_\chi^{\text{LF},m} \right )_{zx}=  
\frac{ e^5 \, v_z^3 \, \tau^2 \, 
\alpha_{J}^{ \frac{2} {J}} \, B_y\, B^2 \,}
	 {128 \, \pi^{\frac{3}{2}}} \,
\frac{ \Gamma  \big( \frac{2 \, J - 1} {J} \big)}
{\,  \Gamma  \big(\frac{ 9 \, J -2 } {2 \, J}\big)
}
\, \tilde{\mathcal{L}}_J^{ m } \,
\Upsilon_{ -\frac{J+2}{J}}(\mu ,T) \,,
\text{ where }
\tilde{\mathcal{L}}^{ m}_J = 
 \, 7 \, J^3+13\, J^2+ J -9 +\frac{2}{J}  \, .
\end{align} 
Here, only the out-of-plane component survives.

\item{$n=2$:}
\\Here, only in-plane components are generated, which turn out to be the same as the expressions shown in Eqs.~\eqref{eqlfxx} and \eqref{eqlfyx}.
This is no surprise because there is no nonzero BC- or OMM-contributed part. The two systems differ only through the value of BC (differing by a factor of two) for the dispersing bands.

\item{$n=3$:}
\\It turns out to be the same as Eq.~\eqref{eqLFtot}, again because there is no nonzero BC- or OMM-contributed part.

\end{enumerate}
Gathering all the contributions shown above, the net $zx$-part evaluates to
\begin{align}
&   \left (\sigma_{ \chi }^{ {\text{LF}}} \right )_{zx}   =
\frac{ -\,
e^3 \, v_z \, J \, \tau^2 \, B_y }
	 {2 \, \pi^2} \,
\Bigg [ \frac{\Upsilon_{1}(\mu ,T) }{3} 
+  
\frac{ e^2 \,  v_z^2 \, B^2 \, \alpha_{J}^{ \frac{2} {J} } \,\sqrt \pi  }
{8}
\, \Bigg \lbrace 
\frac{ \Gamma  \big( \frac{2\, J - 1} {J} \big)}
{8\,\Gamma  \big( \frac{9 \, J - 2} {2\, J} \big)}\, 
 \tilde{\mathcal{L}}_{J,1}
\Upsilon_{ - \frac{ J+2}{J}}(\mu ,T)
 - \frac{\tau^2\, \Gamma  \big( \frac{2\, J - 1} {J} \big)}
 { \Gamma  \big( \frac{7 \, J - 2} {2\, J}
 \big)}\,  
 \tilde{\mathcal{L}}_{J,2} \,
  \Upsilon_{\frac{ J -2}{J}}(\mu ,T)
 \Bigg \rbrace \Bigg ]\, , \nn
&  \text{where }
\tilde{\mathcal{L}}_{J,1} = 5\, J^2-19 \, J - 1 +\frac{9}{J}
-\frac{2}{J^2}
\text{ and } \tilde{\mathcal{L}}_{J,2} = {\mathcal{L}}_{J,2} \,.
\end{align}

\subsubsection{Comparison with RSW semimetals}

Let us compare the $J=1$ TSM case with an \textit{RSW node} (with two valence and two conduction bands), both of which have an isotropic linear-in-$k$ dispersion of the bands. Of course, the $J=1$ TSM corresponds to $v_\perp=v_z$, due to isotropy. In Ref.~\cite{ips-rsw-ph}, although we discussed the results for $n=1$, the results for $n=2$ and $n=3$ were not computed. Due to the presence of two conduction bands for the case of the RSW node, we use the index $\check s$ to label them, where $ \check s \in  \left \lbrace   \frac{1}{2},  \frac{3}{2} \right \rbrace$.
Collecting the results till $n=3$ and upto order $\order{B^3}$, the $\mu>0$ condition leads to
\begin{align}
& \left( \sigma^{\text{LF}, \text{BC}}_{\chi} \right)_{xx}  ({\check s})
= \left( \sigma^{\text{LF}, m }_{\chi} \right)_{xx} ({\check s}) = 0 \,, \quad
\left( \sigma^{\text{LF}, \text{H}}_{\chi} \right)_{xx} ({\check s}) =
\frac{- \,e^4 \,  \tau^3 \, {\check s}^3\, v_z^3 \, \Upsilon_{0} ( \mu, T ) }
{6 \,  \pi^2} \,  B_y^2 \,, \nn
& \left( \sigma^{\text{LF}, \text{BC}}_{\chi} \right)_{yx} ({\check s})
= \left( \sigma^{\text{LF}, m }_{\chi} \right)_{yx} ({\check s})
= 0\,, \quad
\left( \sigma^{\text{LF}, \text{H}}_{\chi} \right)_{yx} ({\check s})
=
\frac{ e^4 \,  \tau^3 \, {\check s}^3\, v_z^3 \,    
\, \Upsilon_{0} ( \mu, T )
}
{6 \,  \pi^2} \, B_x\, B_y\,, \nn
\left( \sigma^{\text{LF}}_{\chi} \right)_{zx}   ({\check s})
& = 
\frac{ -\, e^3 \, {\check s} \, \tau^2 \, v_z \,B_y } 
{ 2 \, \pi^2}  
\left[ 
\frac{ \Upsilon_1(\mu ,T) } {3} 
+ 
\frac{ e^2 \, {\check s}^2 \, v_z^4\,  B^2}  {15}
\left \lbrace \left(
\check{\mathcal{G}}_{\check s}^2 
- 8 \, \check{\mathcal{G}}_{\tilde s} \, {\check s}^2 
+ 3 \, {\check s}^4  \right)  \Upsilon_{-3}(\mu ,T)
-5 \, {\check s}^2\, \tau^2 \,\Upsilon_{-1}(\mu ,T)
\right \rbrace \right],
\end{align}
where $\lbrace 
\check{\mathcal{G}}_{1/2}, \,\check{\mathcal{G}}_{3/2} \rbrace
= \left \lbrace \frac{7}{4}, \, \frac{3}{4} \right \rbrace $. The $zx$-component contains nonzero values of $
\left( \sigma_\chi^{\text{LF}, \text{H}} \right)_{zx}  ({\check s})$, $
\left( \sigma_\chi^{\text{LF}, \text{BC}} \right)_{zx} ({\check s})$, and $ \left( \sigma_\chi^{\text{LF},m} \right)_{zx} ({\check s})$.

\section{Magnetothermoelectric conductivity and magnetothermal coefficient}
 \label{secalpha}

We divide up the expressions for ${\bar \alpha}^{\chi}_{s}  $ and ${\bar \ell}^{\chi}_{s}  $, shown in Eqs.~\eqref{eqbaralpha} and \eqref{eqbarell}, into three parts as
\begin{align}
\label{eqalpha3parts}
{\bar \alpha}^{\chi}_{\tilde s}  =  
\alpha^{\chi, \text{Drude}}_{s} + \alpha^{\chi, \text{BC}}_{\tilde s} 
+ \alpha^{\chi, m}_{s} \text{ and }
{\bar \ell}^{\chi}_{s}  =  
\ell^{\chi, \text{Drude}}_{s} + \ell^{\chi, \text{BC}}_{s} 
+ \ell^{\chi, m}_{s} \,.
\end{align} 
Analogous to the case of ${\bar \sigma}^{\chi}_{s}$, (1) the first part stands for the Drude contribution; (2) the second part arises solely due to the effect of the BC and survives when OMM is set to zero; and (3) the third part is the one which goes to zero if OMM is ignored.
The generic expressions, obtained by expanding Eqs.~\eqref{eqbaralpha} and \eqref{eqbarell} (in powers of $B$), can be found in Ref.~\cite{ips-ruiz}. Here, we only show the final results applicable for TSMs.
Needless to emphasize that, since these in-plane components contain only even powers of $B$, the expressions shown below are correct upto $\order{B^3}$.

\subsection{Longitudinal components}

The part-by-part expressions for the longitudinal components are given by
\begin{align}
\label{eqbaraxx}
& \left (\alpha_\chi^{\text{Drude}} \right )_{xx} = 
 \frac{ -\,e \,  \tau  \,  J \, \mu \,T }
 { 9   \,  v_z} \,,
\quad
\left ( \alpha_\chi^{\text{BC}} \right )_{xx} =
   \frac{ \sqrt{\pi } \, e^3 \,  \tau \, 
   v_{z } \,  \alpha_{J}^{ \frac{2} {J} } \, T}
 { 192 \,   \mu^{ \frac{2+J} { J} } } 
\,\frac{ \Gamma \big ( \frac{2 \, J - 1} {J} \big) }
{ \Gamma \big( \frac{9\, J -2} {2\, J}  \big) } \,    
  \frac{ g_{x}^{bc} (J)\,  B_x^2 
+ g_y^{bc} (J) \,   B_y^2} { J}\,,
\nn & \left (\alpha_\chi^{m} \right )_{xx} =
\frac{ \sqrt{\pi} \, e^3 \,  \tau \, v_z \, \alpha_{J}^{ \frac{2} {J} } \,T}
{ 192 \,  \mu^{\frac{2+J} {J} } }  
\frac{ \Gamma \big ( \frac{2 \, J - 1} {J} \big) }
{ \Gamma \big( \frac{9\, J -2} {2\, J}  \big) } 
 \,  \frac{ g_x^m(J) \,  B_x^2  + g_y^m (J) \,   B_y^2}  { J} \,, 
\end{align}
\begin{align}
\label{eqbarlxx}
& \left ( \ell_\chi^{\text{Drude}} \right )_{xx} 
= \frac{J \, \tau\,\mu^2\, T}
{ 18 \, v_z }  \,,
\quad
(\ell_\chi^{\text{BC}})_{xx}=
\frac{	\sqrt \pi \, e^2\, \tau \, v_z \, 
\alpha_{J}^{ \frac{2} {J} } \, T}
{ 384\,  \mu^{\frac{2} {J}} } 
\,\frac{ \Gamma \big (  \frac{2 \, J - 1} {J} \big) }
{ \Gamma \big( \frac{9\, J -2} {2\, J} \big) }
 \left [  g_x^{bc} (J) \,  B_x^2  
 +   g_y^{bc} (J)\,   B_y^2  \right ] ,
\nn &
\left ( \ell_\chi^{\, m} \right )_{xx} = 
\frac{  \sqrt \pi \,
e^2 \,  \tau \, v_z \,   \alpha_{J}^{ \frac{2} {J} }
\,T  } 
{  384 \, \,\mu^{ \frac{2} {J} }
 } 
\,\frac{ \Gamma \big ( \frac{2 \, J - 1} {J} \big) }
{ \Gamma \big( \frac{9\, J -2} {2\, J}  \big) } 
  \left [  g_x^m (J) \,  B_x^2  +   g_y^{m} (J)\,   B_y^2   \right ] .
\end{align}

\subsection{In-plane transverse components}

The part-by-part expressions for the in-plane transverse components are given by
\begin{align}
\label{eqbarayx}
&  \left ( \alpha_\chi^{\text{Drude}} \right )_{yx} =0 \,, \quad
 \left (\alpha_\chi^{\text{BC}} \right  )_{yx} =
 \frac{ \sqrt{\pi } \, 
  e^3 \,  \tau \,  v_z   \,  \alpha_{J}^{ \frac{2}{J} }\,T  }
  { 96 \,  \mu^{\frac{2+J} {J} }  }  
\,\frac{ \Gamma \big ( \frac{2 \, J - 1} {J} \big) }
{ \Gamma \big( \frac{9\, J -2} {2\, J}  \big) } 
\, \frac{ f^{bc} (J) \, B_x \, B_y} {J}  \, ,
\nn &
\left ( \alpha_\chi^{m} \right )_{yx} = 
\ \frac{ \sqrt{\pi } \, e^3  \,  \tau \, v_z\, 
\alpha_{J}^{ \frac{2} {J} }  \,T}
{ 96 \, \mu^{\frac{2+J} {J} } } 
\frac{ \Gamma \big ( \frac{2 \, J - 1} {J} \big) }
{ \Gamma \big( \frac{9\, J -2} {2\, J}  \big) } 
\, \frac{ f^m ( J) \,  B_x \, B_y} {J}  \,,
\end{align}
\begin{align}
\label{eqbarlyx}
& \left ( \ell_\chi^{\text{Drude}} \right )_{yx} =0\,, \quad
\left ( \ell_\chi^{\text{BC}} \right )_{yx} =
\frac{	\sqrt \pi \, e^2\, \tau \, v_z \, 
\alpha_{J}^{ \frac{2} {J} } \, T}
{ 192 \,   \mu^{\frac{2} {J}} } 
\,\frac{ \Gamma \big (  \frac{2 \, J - 1} {J} \big) }
{ \Gamma \big( \frac{9\, J -2} {2\, J}  \big) }
\, f^{bc} (J)\, B_x \, B_y    \,,\nn &
\left ( \ell_\chi^{\, m} \right )_{yx}  = 
 \frac{	\sqrt \pi \, e^2\, \tau \, v_z \, 
 \alpha_{J}^{ \frac{2} {J} } \, T}
{ 192 \,  \mu^{\frac{2} {J}} } 
\,\frac{ \Gamma \big ( \frac{2 \, J - 1} {J} \big) }
{ \Gamma \big( \frac{9\, J -2} {2\, J}  \big) }
\, f^{m} (J) \, B_x \, B_y   \,.
\end{align}

\begin{figure*}[t]
{\includegraphics[width=0.6\linewidth]{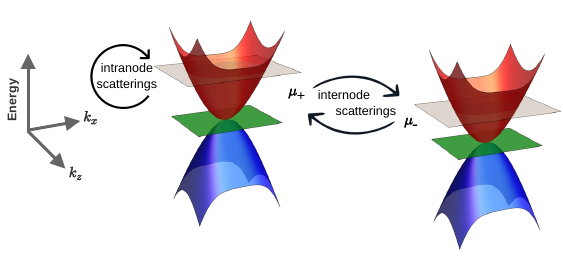}}
\caption{\label{figdis2}Schematics of the scattering processes between two nodes of $J=2$ TSM of opposite chiralities. The values of the chemical potential have been tuned to cut the positive-energy dispersive band at each node.}
\end{figure*}

\subsection{Mott relation and Wiedemann-Franz law}

From the explicit expressions of the in-plane longitudinal and transverse components of $ \bar \sigma_\chi $, $\bar \alpha_\chi $, and $\bar \ell_\chi $ [cf. Eqs.~\eqref{eqbarsxx}, \eqref{eqbarsyx}, \eqref{eqbaraxx}, \eqref{eqbarlxx},\eqref{eqbarayx}, and \eqref{eqbarlyx}], we can immediately verify that the relations,
\begin{align}
\partial_{ \mu }  (\bar \sigma_{\chi}  )_{ij} 
= - \frac  {3\, e }  {\pi^2 \, T} \left (\bar \alpha_{\chi} \right )_{ij} 
+ \order{T^{2}} \text{ and }
	( \bar \sigma_\chi)_{ij} =
\frac{ 3 \,e^2 } {\pi^2\, T}   
\left (\bar \ell_{\chi} \right )_{ij}   + \order{T^{2}},
\label{eqmott}
\end{align}
hold. These two are the Mott relation and the Wiedemann-Franz law, respectively, which relate the three response-tensors in the limit $\beta \rightarrow \infty $ \cite{mermin}. In particular, their validity in the presence of BC and OMM confirms that the relations hold in generic situations \cite{xiao06_berry}.
Consequently, if we know the nature of the magnetoelectric conductivity, we can infer the nature of the remaining two coefficients by using the above equations.

\section{Effects of internode scatterings} 
\label{secinter}

Till now, we have focused only on intranode scatterings, ignoring any internode processes. This section will be dedicated to understanding how the internode scatterings affect the magnetoelectric-conductivity tensor, characterized by a phenomenological relaxation time, $\tau_G $. We will not discuss the corresponding influence on the magnetothermoelectric and magnetothermal coefficients assuming that the Mott relation and the Wiedemann-Franz law will hold, and ensure that their nature can be determined from that of the magnetoelectric-conductivity tensor.

To start with, let us assume that initially, in the infinite past (denoted by time $ t = -\infty $), both the nodes had the same chemical potential $E_F$, characterized by the distribution function
$
f_0 (\varepsilon  ) =\left[ {1 + e^{\frac{\varepsilon - E_F} {T}}} \right ]^{-1}\,,
$
in the absence of any externally applied fields.
Eventually, on applying the electromagnetic fields, there is the onset of the chiral anomaly \cite{adler, bell, chiral_ABJ, hosur-review, son13_chiral, ips-internode}, causing the two conjugate nodes (with $\chi = \pm 1$) to acquire a local equilibrium value of the chemical potential, given by $\mu_\chi$. Therefore, the local equilibrium distribution function at each node is given by
\begin{align}
f^\chi_{s, L} \simeq  f_0 (\xi_{\chi}^{s}) 
 +  \left [- f_{0}^\prime (\xi_{\chi}^{s}) \right ]
 \delta \mu_\chi \,,  \quad
 \delta \mu_\chi \equiv  \mu_\chi - E_F \,.
\end{align}
We define the average of an observable $\mathcal{O}$, at a node with chirality $\chi$, as
\begin{align}
\label{eqaverage}
\bar{\mathcal{O}}_\chi \equiv
\left \langle \mathcal{O}_\chi^s ( \xi_\chi^s (\boldsymbol k), E_F ,T) 
\right  \rangle = 
\frac{
\sum \limits_{ s } 
\int \frac{d^3 \boldsymbol{k}} {(2 \pi)^3} 
\left(  {\mathcal D}_\chi^s (\boldsymbol{k}) \right)^{-1} 
\left [ - f_0^\prime (\xi_\chi^s (\boldsymbol k) ) \right ] 
\mathcal{O}_\chi^s ( \xi_\chi^s (\boldsymbol k), E_F ,T) }
{ \sum \limits_{\tilde s} \int 
\frac{d^3 \boldsymbol{ q} }  {(2 \pi)^3} 
\left( {\mathcal D}_\chi^{\tilde s} (\boldsymbol{q}) \right)^{-1}
\left [ - f_0^\prime (\xi_\chi^{\tilde s} (\boldsymbol{ q}) ) \right ]  } \,,
\end{align}
where $s$ and $\tilde s$ are the band indices.
Let us define
\begin{align}
\mu_G =\frac{\mu_\chi + \mu_{-\chi} } {2} \,,
\quad
\rho_\chi \equiv   \sum \limits_{s} \int 
\frac{d^3 \mathbf{k} }  {(2 \pi)^3} \left( {\mathcal D}_\chi^s (\boldsymbol{k}) \right)^{-1}
\left [ - f_0^\prime (\xi_\chi^{s} (\mathbf{k}) ) \right ],
\quad \rho_G = \frac{ \rho_\chi + \rho_{-\chi} }  {2}\,.
\end{align}
The electric conductivity for the internode-scattering-induced current, whose detailed derivation can be found in Ref.~\cite{ips-internode}, is given by
\begin{align}
\label{eqcond1}
& \left( {\sigma}^{\rm inter }_\chi \right)_{ij} (s) =
\frac{  e^2  \, \rho_{-\chi} }  
{ \rho_\chi \, \rho_G  }  
\left [  \tau_G
-\frac{\tau \, \rho_G } { \rho_{-\chi}}   \right ]
 \int
\frac{ d^3 \boldsymbol k}{(2\, \pi)^3 } 
\left [- f_{0}^\prime (\xi_\chi^s) \right ] 
\;  \left[   \left ( {w}_\chi^s \right)_i
+  \left ( {W}_\chi^s \right)_i  \right]\,
\,  {\mathcal I}_j^\chi    \,,
\nn &
\boldsymbol {\mathcal I}^\chi  = \rho_\chi
\left \langle
\mathcal{D}^{\chi}_{s} 
\left \lbrace {\boldsymbol{w}}^s_\chi 
+ \boldsymbol{W}_\chi^s \right \rbrace\right  \rangle 
=
\sum \limits_s \int \frac{d^3 \boldsymbol{k}} {(2 \pi)^3} 
\left [ - f_0^\prime (\xi^\chi_s (\boldsymbol k) ) \right ] 
 \left \lbrace {\boldsymbol{w}}^s_\chi 
+ \boldsymbol{W}_\chi^s \right \rbrace.
\end{align}

We retain terms upto order $ B^2 $ by using Eq.~\eqref{eqexp}.
The above equation greatly simplifies for the cases when $ \varepsilon_s  $ is a function of magnitudes of the momentum-components (i.e., $ \varepsilon_s  (\boldsymbol k) =  \varepsilon_s  (|k_x|, |k_y|, |k_z|)$). In that situation, $ f_0^{\prime}\big (  \varepsilon_s  (\boldsymbol k) \big) $ and its derivatives (with respect to $ \varepsilon_s  $) are even functions of $\boldsymbol k $. Hence, for our case of TSMs, we need to use
\begin{align}
\label{eqiso}
& \left( {\sigma}^{\rm inter }_\chi \right)_{ij} (s) = 
\frac {e^2  
\left [ \tau_G \, \rho_{-\chi}^{(0)} 
- \tau \, \rho_G^{(0)} \right ]
 \mathcal{Z}_{\chi, i}^s \, \zeta^\chi_j   }
 { \rho_G^{(0)} \, \rho_\chi^{(0)} } 
 + \order{B^3} \,,
\nn & \rho_\chi^{(0)} =
 \sum \limits_{ \tilde s }
\int \frac{ d^3 \boldsymbol q}  {(2\, \pi)^3 } 
 \left \lbrace  - f_0^\prime 
 \big (\varepsilon_{ \tilde s } 
 (\boldsymbol q) \big)  \right \rbrace 
 \Big\vert_{\text{node } \chi} \,,\quad
\rho_G^{(0)} = 
\frac{ \rho_\chi^{(0)} + \rho_{-\chi}^{(0)} }  {2}   \,, 
\quad \zeta^\chi_j  = 
 \sum_{\tilde s}  \mathcal{Z}_{\chi, j }^{\tilde s}  \,,
\nn &  \mathcal{Z}_{\chi, j}^s
=   B_j \int
\frac{ d^3 \boldsymbol k} {(2\, \pi)^3 } 
\left [
 e \, {\boldsymbol \Omega}_\chi^s (\boldsymbol k) 
 \cdot  {\boldsymbol   v}_s(\boldsymbol k) 
\left \lbrace - 
 f_0^\prime \big (\varepsilon_s (\boldsymbol k) \big)  
 \right  \rbrace
+
  \left ( m^s_\chi (\boldsymbol k) \right)_j  
  \left (  v_s (\boldsymbol k) \right)_j
f_0^{\prime \prime} \big (\varepsilon_s (\boldsymbol k)  \big) 
\right ]\Big\vert_{\text{node } \chi} \,.
\end{align}
Ref.~\cite{ips-internode} contains the resulting final expressions when we have the internode scatterings between (1) a single $J=1$ TSM node (at the $ \Gamma $-point) and a double-pseudospin-1/2 node (at the $ R $-point);
(2) a single RSW node (at the $\Gamma$-point) and a $J=1$ double-TSM node (at the $ R $-point).

For the special case where we have scatterings between two nodes of the same nature, with no net energy offset between the nodal points (relative to each other in the BZ), Eq.~\eqref{eqiso} further simplifies to~\cite{ips-internode}
\begin{align}
\label{eqsamenode}
 \left( {\sigma}^{\rm inter }_\chi \right)_{ij}  (s)
 & =
 \frac { e^2   \left (\tau_G - \tau \right) }
{ \rho_1^{(0)} } \,
 \mathcal{Z}^{s}_{1,i} \, \sum \limits_{\tilde s} \mathcal{Z}^{\tilde s}_{1,j}\,.
\end{align}
Here, we consider this situation involving the single band with $s=1$ at each node of TSM, as shown in Fig.~\ref{figdis2}.
Therefore, for our set-up, the nonzero components evaluate to
\begin{align}
 \left( {\sigma}^{\rm inter }_\chi \right)_{xx} 
 & = \frac { 49 \, e^4\, v_z \,J^3 \, \alpha_J^{\frac{2}{J}}
 \left (\tau_G - \tau \right)}
{ 36 \, \pi^\frac{5}{2} \, E_F^{\frac{2}{J}}} \,
 \frac{  \Gamma  \big( \frac{J+2} {2\, J}   \big) }
{  \Gamma \big (  \frac{1}{J} \big)} B_x^2\,, \quad
 \left( {\sigma}^{\rm inter }_\chi \right)_{yx}  
 = \frac { 49 \, e^4\, v_z \,J^3 \, \alpha_J^{\frac{2}{J}}
 \left (\tau_G - \tau \right)}
{ 36 \, \pi^\frac{5}{2} \, E_F^{\frac{2}{J}}} \,
 \frac{  \Gamma  \big( \frac{J+2} {2\, J}  \big) }
{  \Gamma \big (  \frac{1}{J} \big)}\,   B_x \, B_y \,.
\end{align}
Side by side, we invoke the analogous scenario with two conjugate nodes of the WSM/mWSM type, which gives us
\begin{align}
 \left( {\sigma}^{\rm inter }_\chi \right)_{xx}   \Big \vert_{\text{mWSM}}
 & = \frac { 4 \, e^4\, v_z \,J^3 \, \alpha_J^{\frac{2}{J}}
 \left (\tau_G - \tau \right)}
{ 9 \, \pi^\frac{5}{2} \, E_F^{\frac{2}{J}}} \,
 \frac{  \Gamma  \big( \frac{J+2} {2\, J}   \big) }
{  \Gamma \big (  \frac{1}{J} \big)} B_x^2\,, \quad
 \left( {\sigma}^{\rm inter }_\chi \right)_{yx}  \Big \vert_{\text{mWSM}}
 = \frac { 4 \, e^4\, v_z \,J^3 \, \alpha_J^{\frac{2}{J}}
 \left (\tau_G - \tau \right)}
{ 9 \, \pi^\frac{5}{2} \, E_F^{\frac{2}{J}}} \,
 \frac{  \Gamma  \big( \frac{J+2} {2\, J}  \big) }
{  \Gamma \big (  \frac{1}{J} \big)}\,   B_x \, B_y \,.
\end{align}

\section{Summary, discussions, and future perspectives}
\label{secsum}

As a follow-up of our investigations of transport signatures for twofold and multifold 3d semimetals, considering planar-Hall and planar-thermal Hall configurations, we have studied here nodes hosting pseudospin-1 quasiparticles. Resorting to the semiclassical Boltzmann equations and relaxation-time approximations, we have chalked out all the nonzero components of the linear-response tensors, assuming the weak (i.e., nonquantizing) magnetic-field limit. In order to have a complete description, we have incorporated the effects of both the Berry curvature and the orbital magnetic moment, since both of these arise from the underlying topological features of the bandstructure. Going beyond our previous works, we have computed the out-of-plane component of the electric conductivity, which are caused by the intrinsic anomalous-Hall and the Lorentz-force-contributed currents. It is evident from our results that the Lorentz-force operator gives rise to in-plane components, in addition to the out-of-plane ones. Last but not the least, we have worked out the response characteristics induced by internode scatterings. Our theoretical explorations involving TSMs are particularly important in the context of contemporary experiments, e.g., the one reported in Ref.~\cite{claudia-multifold}. In fact, we have elucidated the results upto $\order{B^3}$, because the data-fitting in Ref.~\cite{claudia-multifold} has been implemented via a phenomenological model (describing the $J=1$ TSM) comprising terms upto order $B^3$. Hence, the relevance and timeliness of our studies cannot be overemphasized.

The results for the in-plane components bear qualitative resemblance to those discussed in Ref.~\cite{ips-ruiz}, which deals with WSMs and mWSMs (having twofold band-crossings). This is because, at the zeroth order in the OMM, the flat-band does not contribute to transport --- hence, we have omitted their contributions, if any. The $J=1$ case also resembles the RSW case, studied in Ref.~\cite{ips-rsw-ph}, because both the systems are isotropic and have linear-in-momentum dispersions. In fact, here we have derived and compared some higher-order-in-$\check L$ terms of an RSW node, in order to make a clear comparison between the characteristics for the $J=1 $ TSM case and the RSW case.

As emphasized above, we have here employed the relaxation-time approximations, both for the intranode- and internode-scattering processes. Hence, it remains to be seen if any new information is unravelled
by going beyond the relaxation-time approximations \cite{timm, girsh-internode-spin1}.
There are multiple other directions that can be pursued in the context of the studies reported here. One of those involves repeating our calculations in the presence of nonzero tilts of the TSM nodes \cite{timm, das-agarwal_omm, rahul-jpcm, ips-tilted, ips-shreya, ips_tilted_dirac}. In particular, tilting causes linear-in-$B$ terms to appear in the in-plane response coefficients, as found in Refs.~\cite{rahul-jpcm, ips-tilted, das-agarwal_omm, ips-tilted, ips-shreya, ips_tilted_dirac}. In this connection, it is worth mentioning that a chiral pseudomagnetic field, induced by elastic deformations, can also give rise to $B$-linear terms, as elaborated on in earlier works \cite{onofre, ips_rahul_ph_strain, ips-ruiz, ips-rsw-ph}.
Next, it will be worthwhile to study the transport properties under a strong quantizing magnetic field, when we will need to consider the energy levels being quantized into discrete Landau levels \cite{ips-kush, fu22_thermoelectric, staalhammar20_magneto, yadav23_magneto}. Lastly, in order to simulate more realistic scenarios, we need to consider the effects of disorder and/or many-body interactions. This will necessitate incorporating state-of-the-art many-body formalisms \cite{ips-seb, ips_cpge, ips-biref, ips-klaus, rahul-sid, ipsita-rahul-qbt, ips-qbt-sc, ips-hermann-review}. 

\section*{Acknowledgments}
We thank Rahul Ghosh for useful discussions.

\appendix

\section{Identities for some useful integrals} 

\label{appint}

In order to solve the various conductivity expressions, we have to perform integrals of the form
\begin{align}
\mathcal{I} = \int \frac{d^{3} {\boldsymbol{k}} }
{(2 \pi )^{3}} \,\mathcal{T} ({\boldsymbol{k}} , \xi_\chi^s ) \,
 f_{0}^\prime (\xi_\chi^s)  \,,
\end{align}
where $\xi_\chi^s =  \varepsilon_s (\boldsymbol{k}) + \varepsilon_\chi^{ (m) }(\boldsymbol{k}) $.
Here, we focus only the cases when the checmical potential ($\mu$) cuts the positive-energy band(s) and, therefore, $\varepsilon_s (\boldsymbol{k}) > 0$.
Observing that the cylindrical symmetry of the system can be utilized to evaluate the integrals, we employ the following coordinate transformation:
\begin{align}
\label{eqcyln} 
k_x =  \left( \frac{ \varepsilon}{\alpha_{J} } \sin \gamma \right)^{1/J} \cos \phi \,, \quad
k_y =  \left( \frac{ \varepsilon}{\alpha_{J} } \sin \gamma \right)^{1/J} \sin \phi\,, 
\quad k_{z} = \frac{ \varepsilon}{v_z } \cos \gamma\,,
\end{align}
where $ \varepsilon \in [0, \infty )$, $\phi \in [0, 2 \pi )$, and $\gamma \in [0, \pi ]$. 
The Jacobian for the transformation is
$\mathcal{J} ( \varepsilon , \gamma ) =   \frac{1} {J\, v_z \sin \gamma } 
\left( \frac{ \varepsilon \sin \gamma}
{\alpha_{J} } \right) ^{2/J}$. 
We rewrite the integral using the following substitutions:
\begin{align}
\int_{- \infty}^{ \infty} d^3\boldsymbol{k }  
\rightarrow   \int_{- \infty}^{ \infty} d \varepsilon  
\int_{0}^{ 2 \pi} d \phi \,   \int_{0}^{  \pi} d \gamma \,   \mathcal{J} ( \varepsilon , \gamma ) \;\;\;\;
\text{ and } \;\;\;\;
\xi_\chi^s(\boldsymbol{k})   \rightarrow  
\xi_\chi^s (\varepsilon)   
=    \varepsilon + \varepsilon_\chi ^{ (m) } \,.
\end{align}
With the implementation of the above coordinate transformation, the original integral evolves into
\begin{align}
\mathcal I & = \frac{1}{(2 \pi )^{3}}  \int_{- \infty}^{ \infty} d \varepsilon  
\int_{0}^{ 2 \pi} d \phi \,   \int_{0}^{  \pi} d \gamma \, 
\mathcal{H} (  \varepsilon, \phi , \gamma , \xi_\chi^s )   \, 
 f^\prime_{0} (\xi_ s ^\chi) 
\quad \left[ \text{where } \mathcal{H} (  \varepsilon, \phi , \gamma , \xi_\chi^s )
=  \mathcal{J} ( \varepsilon , \gamma ) \, 
 \mathcal{T}  (  \varepsilon, \phi , \gamma , \xi_\chi^s ) \right]
\nn & = \frac{1}{(2 \pi )^{3}}  \int_{0}^{\infty} d \varepsilon \,   
\mathcal{K} (\chi,  \varepsilon )   \,  f^\prime_{0} (\xi_\chi^s) 
\quad \left[ \text{where }
\mathcal{K} ( \chi,  \varepsilon  )
 = \int_{0}^{2 \, \pi} d \phi \,  \int_{0}^{\pi} d \gamma \, 
 \mathcal{H} (  \varepsilon, \phi , \gamma , \xi_\chi^s ) \right ] .
\end{align} 
For the $ \varepsilon$-integration, we apply the Sommerfeld expansion \cite{mermin}, which is valid under the condition $ 1/(\beta \, \mu) \ll 1$. This implies that we use the identity
\begin{align}
\label{eqsom}
\int_{0}^{ \infty} d \varepsilon   \,   
  \varepsilon^{n}   \left [ - f^\prime_{0} (  \varepsilon )\right ] 
& =  \Upsilon_{n} (\mu ) 
=  \mu^{n} \,  \left[ 1 + \frac{\pi^2 \,  n \,  (n-1)}
{6 \left(  \beta \,  \mu \right)^{2} } 
+ \order{\left( {\beta \,\mu}\right)^{-3}} \right] ,
\end{align}
where $\beta (\boldsymbol{r})= 1/T (\boldsymbol{r})$.

For integrals involving higher-order derivatives of $f_0$, we have 
\begin{align}
\int_{0}^{ \infty} d \varepsilon    \,    \varepsilon ^{n}  \,(-1)^{{ \lambda }+1}  \,
\frac{\partial^{{ \lambda }+1} 
	\, f_{0} (  \varepsilon  ) } { \partial  \varepsilon ^{{ \lambda }+1} }  
= \frac{n!}{(n-{ \lambda })!}  \, \Upsilon_{n-{ \lambda }} (\mu )\,. 
\label{Identity3}
\end{align}

For the thermoelectric and thermal tensors, we need to use the identity 
\begin{align}
\int_{0}^{ \infty} d \varepsilon   \,  \varepsilon^{n}  \,  ( \varepsilon - \mu ) 
\,  (-1)^{{ \lambda }+1}  
\, \frac{\partial^{{ \lambda }+1} f_{0} (  \varepsilon ) }
{ \partial  \varepsilon^{{ \lambda }+1} } 
= \frac{(n+1)!}{(n+1-{ \lambda })!}  \, 
\Upsilon_{n+1-{ \lambda }} (\mu )- \mu \, \frac{n!}{(n-{ \lambda })!}  
\, \Upsilon_{n-{ \lambda }} (\mu ) \,. 
\label{Identity4}
\end{align}

\section{Current from the Lorentz-force part}
\label{appLF}

 In this appendix, we deal with the so-called Lorentz-force part, which arises from the current density of (see Refs. \cite{ips-rsw-ph, ips_tilted_dirac} for a detailed derivation)
 \begin{align} 
& {\boldsymbol J}^{s, \rm LF}_\chi = 
-\,e^2 \,  \tau \int \frac{d^3 \boldsymbol{k}} {(2 \, \pi)^3} 
\left(  {{\boldsymbol w}}_\chi^s 
+   {{\boldsymbol W}}_\chi^s \right )
\, f_{0}^\prime (\xi_\chi^s)\, \mathcal{Y}_\chi^s ,
  \text{ where } \check{L} 
  = ({\boldsymbol w}_\chi^s \cross \boldsymbol{B}) \cdot \nabla_{\boldsymbol{k}}\,,
\nn &
\boldsymbol{W}_\chi^s  
= e \left  ( {\boldsymbol w}_\chi^s \cdot 
  \boldsymbol {\Omega}_\chi^s \right  ) \boldsymbol{B}\,,
  \quad
\mathcal{Y}_\chi^s =  
\sum_{n = 1}^{\infty}
\left (e \, \tau \, \mathcal{D}_\chi^s \right )^n \check{L}^n 
\left [    \mathcal{D}_\chi^s \left
\lbrace {{\boldsymbol w}}_\chi^s 
+  {{\boldsymbol W}}_\chi^s  \right 
\rbrace \cdot \boldsymbol{E}   \right ].  
\end{align} 
This part arises from the action of the Lorentz-force operator $\check{L} $.
The nomenclature reflects the fact that it includes the classical Hall effect due to the Lorentz force. The corresponding components of the electric conductivity is
 \begin{align} 
\left(\sigma_\chi^{\rm LF} \right)_{i j} (s) & = 
-\,e^2 \,  \tau \int \frac{d^3 \boldsymbol{k}} {(2 \, \pi)^3} 
\, \left[  (w_\chi^s)_i  + (W_\chi^s)_i \right ]
\, f_{0}^\prime (\xi_\chi^s) \,
\frac{\partial \mathcal{Y}_\chi^s } {\partial E_j}\,.
\end{align}
The solution is obtained by taking the terms in the summation upto a chosen value of $n$ and, thereafter, expanding the expressions upto the desired power in $B$.

\subsection{$n=1$: Terms originating from the linear action of the Lorentz-force operator} 
\label{appn1}

The $n=1$ term leads to the current density of
 \begin{align} 
& {\boldsymbol J}^{s, \rm LF}_\chi = 
-\,e^3 \,  \tau^2 \int \frac{d^3 \boldsymbol{k}} {(2 \, \pi)^3} 
\, \left[ {{\boldsymbol w}}_\chi^s 
+   {{\boldsymbol W}}_\chi^s \right]
\, \mathcal{D}_\chi^s
\, f_{0}^\prime (\xi_\chi^s)
\left( t_1+ t_2   \right ), \nn
& t_1= \mathcal{D}_\chi^s \,
\check{L}
\left [    \left
\lbrace {\boldsymbol w}_\chi^s 
+ {\boldsymbol W}_\chi^s \right 
\rbrace \cdot \boldsymbol{E}   \right ] ,\quad
t_2 =\left [   \left
\lbrace {{\boldsymbol w}}_\chi^s 
+ {{\boldsymbol W}}_\chi^s \right 
\rbrace \cdot \boldsymbol{E}   \right ]
\check{L} \,
\mathcal{D}_\chi^s\,.
\end{align}
Expanding the integrand upto $\order{ B^3}$, we obtain
\begin{align}
 t_1 &=
 \lbrace 1 -e \left (\boldsymbol{\Omega }_\chi^s \cdot   \boldsymbol{B} \right)
 + e^2   \left (\boldsymbol{\Omega }_\chi^s \cdot   \boldsymbol{B}  \right)^2 \rbrace 
  ({\boldsymbol v}_s \cross \boldsymbol{B}) \cdot \nabla_{\boldsymbol{k}}
  \left( {\boldsymbol v}_s \cdot \boldsymbol{E} \right )
+ ({\boldsymbol v}_s \cross \boldsymbol{B})
 \cdot \nabla_{\boldsymbol{k}} 
 \left( {\boldsymbol  U}^{(m)}_\chi   \cdot \boldsymbol{E} \right ) \nn
 & \qquad  +\lbrace 1 -e \left (\boldsymbol{\Omega }_\chi^s 
 \cdot   \boldsymbol{B}  \right) \rbrace 
  ({\boldsymbol v}_s \cross \boldsymbol{B}) \cdot 
  \nabla_{\boldsymbol{k}}
  \left[ \left( \boldsymbol {u^{(m)}_\chi  + V}_\chi^s \right) \cdot \boldsymbol{E} \right]
+ ({\boldsymbol  u}^{(m)}_\chi  \cross \boldsymbol{B}) \cdot 
\nabla_{\boldsymbol{k}}\, \left[ \left( {\boldsymbol  u}^{(m)}_\chi
 + {\boldsymbol V}^s_\chi \right) \cdot \boldsymbol{E} \right]  \nn
 & \qquad 
 +\lbrace 1 -e \left (\boldsymbol{\Omega }_\chi^s \cdot   
 \boldsymbol{B}  \right) \rbrace 
  ({\boldsymbol  u}^{(m)}_\chi  \cross \boldsymbol{B}) \cdot \nabla_{\boldsymbol{k}} 
   \left(  {\boldsymbol  v}_s  \cdot \boldsymbol{E} \right ),
\end{align}
and
\begin{align} 
t_2 &=
\left ( {\boldsymbol v}_s \cdot \boldsymbol{E} \right)
 ({\boldsymbol v}_s \cross \boldsymbol{B})  \cdot \nabla_{\boldsymbol{k}}
 \left[  -e \left (\boldsymbol{\Omega }_\chi^s \cdot   \boldsymbol{B}  \right)
 + e^2   \left (\boldsymbol{\Omega }_\chi^s \cdot   \boldsymbol{B}  \right)^2 \right]
 + \left ( {\boldsymbol v}_s \cdot \boldsymbol{E} \right)
 ({\boldsymbol  u}^{(m)}_\chi  \cross \boldsymbol{B})  \cdot \nabla_{\boldsymbol{k}}
 \left[  -e \left (\boldsymbol{\Omega }_\chi^s \cdot   \boldsymbol{B}  \right) \right] \nn
 & \qquad 
 +  \left ( {\boldsymbol u}_\chi^{(m)} + {\boldsymbol V}_\chi^s\right)\cdot \boldsymbol{E} 
 \left({\boldsymbol v}_s \cross \boldsymbol{B}\right)  \cdot \nabla_{\boldsymbol{k}}
 \left[  -e \left (\boldsymbol{\Omega }_\chi^s \cdot   \boldsymbol{B}  \right) \right] .
\end{align}

Let us express the current density as
 \begin{align} 
& {\boldsymbol J}^{s, \rm LF}_\chi = 
-e^3 \,  \tau^2 \int \frac{d^3 \boldsymbol{k}} {(2 \, \pi)^3}  
\sum_{\delta =1}^3 
\boldsymbol {\mathcal N}_{1,\delta} \,,
\end{align}
where $\boldsymbol {\mathcal N}_{1, \delta} $ has a $B^\delta $-dependence.
These evaluate to the following expressions:
\begin{enumerate}

\item Linear-in-$B$:
\begin{align}
\boldsymbol {\mathcal N}_{1,1} &=
   {\boldsymbol v}_s\,  f^\prime_0  (\varepsilon_s )
   \left({\boldsymbol v}_s \cross \boldsymbol{B}\right)  \cdot \nabla_{\boldsymbol{k}} 
   \left ( {\boldsymbol v}_s \cdot \boldsymbol{E} \right) .
\end{align}

\item
Quadratic-in-$B$:
\begin{align}
\boldsymbol {\mathcal N}_{1,2}  &={\boldsymbol v}_s 
   \left ( {\boldsymbol v}_s \cdot \boldsymbol{E} \right) f^\prime_0  (\varepsilon_s ) 
    ({\boldsymbol v}_s \cross \boldsymbol{B})  \cdot \nabla_{\boldsymbol{k}}  
  \left[
-e \left (\boldsymbol{\Omega }_\chi^s \cdot  \boldsymbol{B}  \right) \right] 
   + {\boldsymbol v}_s \,  f^\prime_0  \left(\varepsilon_s  \right)
     \left({\boldsymbol v}_s \cross \boldsymbol{B}\right)  \cdot \nabla_{\boldsymbol{k}} 
     \left[  \left(
     {\boldsymbol  u}^{(m)}_\chi   + {\boldsymbol V}_\chi^s \right) \cdot \boldsymbol{E} \right] 
   \nn
   &\quad
   +\left[ \left \lbrace -2 \, e \left( \boldsymbol{\Omega }_\chi^s \cdot  \boldsymbol{B}   \right) {\boldsymbol v}_s   
   + \left(  {\boldsymbol  u}^{(m)}_\chi +  {\boldsymbol V}_\chi^s \right) \right \rbrace 
   f^\prime_0  (\varepsilon_s )
   - \left( \boldsymbol {m}_\chi^s \cdot \boldsymbol{B}   \right) {\boldsymbol v}_s 
   \,   f^{\prime \prime}_0  \left( \varepsilon_s  \right) 
\right]
\left( {\boldsymbol v}_s \cross \boldsymbol{B} \right)  \cdot \nabla_{\boldsymbol{k}} 
\left ({\boldsymbol v}_s \cdot  \boldsymbol{E}   \right ) \nn
   & \quad +{\boldsymbol v}_s \, f^{ \prime}_0  (\varepsilon_s ) 
   \left( {\boldsymbol  u}^{(m)}_\chi  \cross \boldsymbol{B} \right)  \cdot \nabla_{\boldsymbol{k}} 
    \left( {\boldsymbol v}_s \cdot  \boldsymbol{E}   \right ).
 \end{align} 

\item
Cubic-in-$B$:
\begin{align}
\boldsymbol {\mathcal N}_{1,3}  &=
\Big[
\left \lbrace -e \left(  \boldsymbol{\Omega}_\chi^s 
\cdot  \boldsymbol{B}\right)  {\boldsymbol v}_s 
\left( {\boldsymbol v}_s \cdot  \boldsymbol{E}  \right)
+ {\boldsymbol  v}_s  \left( {\boldsymbol  u}^{(m)}_\chi  +{\boldsymbol V}_\chi^s   \right)
 \cdot  \boldsymbol{E} 
  + \left( {\boldsymbol v}_s \cdot  \boldsymbol{E}  \right)
     \left( {\boldsymbol  u}^{(m)}_\chi  +{\boldsymbol V}_\chi^s   \right)
  \right \rbrace
      f^{ \prime}_0  (\varepsilon_s )
 - \left( \boldsymbol {m}_\chi^s \cdot \boldsymbol{B}   \right) {\boldsymbol v}_s
       \left( {\boldsymbol v}_s \cdot  \boldsymbol{E}  \right)
  f^{\prime \prime}_0  (\varepsilon_ s) \Big]      \nn
& \qquad \times
 \left[   \left({\boldsymbol v}_s \cross \boldsymbol{B}\right) 
    \cdot \nabla_{\boldsymbol{k}}
\left \lbrace -e \left( \boldsymbol{\Omega }_\chi^s 
\cdot  \boldsymbol{B}  \right)\right \rbrace \right ] \nn
 & \quad
 + {\boldsymbol v}_s  \left( {\boldsymbol v}_s \cdot  \boldsymbol{E}  \right) 
 f^{ \prime}_0  (\varepsilon_s )
     \left[ \left({\boldsymbol v}_s \cross \boldsymbol{B}\right) 
    \cdot \nabla_{\boldsymbol{k}}
    \left[ e^2 \left( \boldsymbol{\Omega }_\chi^s \cdot  \boldsymbol{B}    
  \right)^2 \right]
    + \left({\boldsymbol  u}^{(m)}_\chi  \cross \boldsymbol{B}\right)
     \cdot \nabla_{\boldsymbol{k}}
    \left \lbrace -e \left( \boldsymbol{\Omega }_\chi^s \cdot  \boldsymbol{B} 
  \right) \right \rbrace  \right] \nn
 & \quad  +
     \Big[ \left \lbrace 3\, e^2\left( \boldsymbol{\Omega }_\chi^s 
  \cdot  \boldsymbol{B}    \right)^2  {\boldsymbol v}_s
     -2\, e \left( \boldsymbol{\Omega }_\chi^s \cdot  \boldsymbol{B} 
     \right)  \left({\boldsymbol  u}^{(m)}_\chi +{\boldsymbol V}_\chi^s   \right)
     + {\boldsymbol  U}^{(m)}_\chi \right \rbrace
      f^{\prime}_0  (\varepsilon_s )
 \nn & \hspace{ 1 cm }      
 + \Big\{ 2\, e \left( \boldsymbol{\Omega }_\chi^s \cdot  \boldsymbol{B}  \right)  {\boldsymbol v}_s 
- \left({\boldsymbol  u}^{(m)}_\chi +{\boldsymbol V}_\chi^s   \right)
       \Big\}   \left( \boldsymbol{m}_\chi^s \cdot  \boldsymbol{B} \right)
        f^{ \prime \prime}_0  (\varepsilon_s )
     + \frac{ \left( \boldsymbol{m}_\chi^s \cdot  \boldsymbol{B} \right)^2}{2}
\, {\boldsymbol v}_s\, 
         f^{ \prime \prime \prime}_0  (\varepsilon_s )
  \Big] 
      \left({\boldsymbol v}_s \cross \boldsymbol{B}\right)  \cdot
   \nabla_{\boldsymbol{k}} \left ( {\boldsymbol v}_s \cdot \boldsymbol{E} \right)\nn
& \quad +  
 \left[ \left \lbrace-2\, e \left( \boldsymbol{\Omega }_\chi^s \cdot  \boldsymbol{B}    \right) 
      {\boldsymbol v}_s 
      + \left({\boldsymbol  u}^{(m)}_\chi  + {\boldsymbol V}_\chi^s    \right) \right \rbrace  
      f^{ \prime }_0  (\varepsilon_s )
      -\left( \boldsymbol{m}_\chi^s \cdot  \boldsymbol{B} \right) {\boldsymbol v}_s 
 \, f^{ \prime \prime}_0  (\varepsilon_s )
      \right]  \nn
& \qquad \times
\Big[ ({\boldsymbol v}_s \cross \boldsymbol{B})  
   \cdot \nabla_{\boldsymbol{k}} 
     \left \lbrace  
 \left ({\boldsymbol  u}^{(m)}_\chi   + {\boldsymbol V}_\chi^s \right) 
 \cdot \boldsymbol{E}  \right \rbrace  
 +  ({\boldsymbol  u}^{(m)}_\chi  \cross \boldsymbol{B})  
   \cdot \nabla_{\boldsymbol{k}} 
     \left (  {\boldsymbol v}_s \cdot \boldsymbol{E} \right )
    \Big]  
\nn  & \quad
+ {\boldsymbol v}_s \,f^{ \prime}_0  (\varepsilon_s ) 
    \left[ ({\boldsymbol v}_s \cross \boldsymbol{B})
     \cdot \nabla_{\boldsymbol{k}}
\left (  {\boldsymbol  U}^{(m)}_\chi  \cdot \boldsymbol{E} \right )
+  \left({\boldsymbol  u}^{(m)}_\chi  \cross \boldsymbol{B}\right)
     \cdot \nabla_{\boldsymbol{k}} 
 \left \lbrace 
 \left  ({\boldsymbol  u}^{(m)}_\chi   
 + {\boldsymbol V}_\chi^s \right) \cdot 
 \boldsymbol{E} \right \rbrace  \right].
\end{align}

\end{enumerate}

\subsection{$n=2$: Terms originating from the quadratic action of the Lorentz-force operator} 
\label{appn2}

The $n=2$ term leads to the current density of
 \begin{align} 
{\boldsymbol J}^{s, \rm LF}_\chi = 
-e^4 \,  \tau^3 \int \frac{d^3 \boldsymbol{k}} {(2 \, \pi)^3} 
 \lbrace {{\boldsymbol w}}_\chi^s 
+   {{\boldsymbol W}}_\chi^s \rbrace
 \left( \mathcal{D}_\chi^s \right)^2
 f_{0}^\prime (\xi_\chi^s)\,
\check{L}^2
\left [  \mathcal{D}_\chi^s \left
\lbrace {{\boldsymbol w}}_\chi^s 
+{{\boldsymbol W}}_\chi^s  \right 
\rbrace \cdot \boldsymbol{E}   \right ] .
\end{align}
Due to the presence of $\check{L}^2$, there is no linear-in-$B$ term here. Let us express the current density as
 \begin{align} 
& {\boldsymbol J}^{s, \rm LF}_\chi = 
-\,e^4 \,  \tau^3 \int \frac{d^3 \boldsymbol{k}} {(2 \, \pi)^3}  
\sum_{\delta =2}^3 \boldsymbol {\mathcal N}_{2,\delta} \,,
\end{align}
where $\boldsymbol {\mathcal N}_{2, \delta} $ has a $B^\delta $-dependence.
These evaluate to the following expressions:
\begin{enumerate}

\item Quadratic-in-$B$:
\begin{align}
\label{eqb12}
\boldsymbol {\mathcal N}_{2, 2} &=
   {\boldsymbol v}_s\, f^\prime_0  (\varepsilon_s ) \left({\boldsymbol v}_s \cross \boldsymbol{B}\right) 
   \cdot \nabla_{\boldsymbol{k}}
   \left[
   \left({\boldsymbol v}_s \cross \boldsymbol{B}\right)  \cdot \nabla_{\boldsymbol{k}} 
   \left ( {\boldsymbol v}_s \cdot \boldsymbol{E} \right)\right].
\end{align}

\item 
Cubic-in-$B$:
\begin{align}
\boldsymbol {\mathcal N}_{2, 3} &=
     {\boldsymbol v}_s\, f^\prime_0  (\varepsilon_s )\;
\left({\boldsymbol v}_s \cross \boldsymbol{B}\right)  \cdot     
   \nabla_{\boldsymbol{k}} 
 \Big[  
 -e \left(  \boldsymbol{\Omega }_\chi^s \cdot  \boldsymbol{B}\right) 
     \left( {\boldsymbol v}_s \cross  \boldsymbol{B}  \right)
     \cdot \nabla_{\boldsymbol{k}} \left ( {\boldsymbol v}_s \cdot \boldsymbol{E} \right)
+  \left( {\boldsymbol v}_s \cross  \boldsymbol{B}  \right) 
     \cdot \nabla_{\boldsymbol{k}} \left\lbrace \left ( {\boldsymbol u}_\chi^{(m)} 
  +  {\boldsymbol V}_\chi^s \right)  
     \cdot \boldsymbol{E} \right \rbrace
 \nn & \hspace{ 4.5 cm}
+ \left( {\boldsymbol  u}^{(m)}_\chi  \cross  \boldsymbol{B}  \right) \cdot 
      \nabla_{\boldsymbol{k}} \left ( {\boldsymbol v}_s \cdot \boldsymbol{E} \right) 
\Big]   
\nn & \quad
+  {\boldsymbol v}_s\, f^\prime_0  (\varepsilon_s )
       \left( {\boldsymbol  u}^{(m)}_\chi  \cross  \boldsymbol{B}  \right) \cdot 
      \nabla_{\boldsymbol{k}} \left[ \left( {\boldsymbol v}_s \cross  \boldsymbol{B}  \right) 
   \cdot \nabla_{\boldsymbol{k}} 
   \left( {\boldsymbol v}_s \cdot \boldsymbol{E} \right)  \right] 
\nn & \quad
 + {\boldsymbol v}_s\, f^\prime_0  (\varepsilon_s )
       \left( {\boldsymbol v}_s \cross  \boldsymbol{B}  \right) \cdot 
      \nabla_{\boldsymbol{k}} \left[ \left( {\boldsymbol v}_s \cdot \boldsymbol{E}  \right) 
      \left( {\boldsymbol v}_s \cross  \boldsymbol{B}  \right) \cdot 
      \nabla_{\boldsymbol{k}} \left \lbrace - e \left (  \boldsymbol{\Omega }_\chi^s \cdot 
      \boldsymbol{B} \right) \right \rbrace \right]\nn
& \quad  
+\Big[ \left \lbrace -2\, e\, {\boldsymbol v}_s \left (  \boldsymbol{\Omega }_\chi^s 
\cdot \boldsymbol{B} \right) + 
{\boldsymbol  u}^{(m)}_\chi 
+ {\boldsymbol V}_\chi^s \right \rbrace 
f^\prime_0  (\varepsilon_s )
-\left( \boldsymbol {m}_\chi^s \cdot \boldsymbol{B}   \right) {\boldsymbol v}_s \,  
 f^{\prime \prime}_0  (\varepsilon_s )\Big]
     \left({\boldsymbol v}_s \cross \boldsymbol{B}\right)
\cdot  
\left [\nabla_{\boldsymbol{k}}
  \left \lbrace  
\left( {\boldsymbol v}_s \cross  \boldsymbol{B}  \right) \cdot 
 \nabla_{\boldsymbol{k}} 
 \left (  {\boldsymbol v}_s \cdot \boldsymbol{E} \right) 
  \right \rbrace \right ].
\end{align}

\end{enumerate}

\subsection{$n=3$: Terms originating from the cubic action of the Lorentz-force operator} 
\label{appn3}

The $n=3 $ term leads to the current density of
 \begin{align} 
{\boldsymbol J}^{s, \rm LF}_\chi = 
-e^5 \,  \tau^4 \int \frac{d^3 \boldsymbol{k}} {(2 \, \pi)^3} 
\lbrace {{\boldsymbol w}}_\chi^s 
+   {{\boldsymbol W}}_\chi^s \rbrace
 \left( \mathcal{D}_\chi^s \right)^3
 f_{0}^\prime (\xi_\chi^s)\, 
\check{L}^3
\left [    \mathcal{D}_\chi^s \left
\lbrace {{\boldsymbol w}}_\chi^s 
+{{\boldsymbol W}}_\chi^s  \right 
\rbrace \cdot \boldsymbol{E}   \right ] .
\end{align}
Due to the presence of $\check{L}^3 $, only a cubic-in-$B$ term needs to be extracted here. 
We can express the current density as
 \begin{align} 
& {\boldsymbol J}^{s, \rm LF}_\chi = -\, e^5 \,  \tau^4
 \int \frac{d^3 \boldsymbol{k}} {(2 \, \pi)^3} \, 
 \boldsymbol {\mathcal N}_{3,3} \,,
 \text{ where }
 \boldsymbol {\mathcal N}_{3, 3}  &=
   {\boldsymbol v}_s\, f^\prime_0  (\varepsilon_s ) 
  \left({\boldsymbol v}_s \cross \boldsymbol{B}\right) 
   \cdot \nabla_{\boldsymbol{k}} \left[
   \left({\boldsymbol v}_s \cross \boldsymbol{B}\right)  \cdot \nabla_{\boldsymbol{k}}
   \lbrace \left({\boldsymbol v}_s \cross \boldsymbol{B}\right)  
   \cdot \nabla_{\boldsymbol{k}}
   \left ( {\boldsymbol v}_s \cdot \boldsymbol{E} \right)\rbrace \right].
   \label{eqb15}
\end{align}

 
\bibliography{ref_spin1}


\end{document}